\documentclass[conference]{IEEEtran}
\def\BibTeX{{\rm B\kern-.05em{\sc i\kern-.025em b}\kern-.08em
    T\kern-.1667em\lower.7ex\hbox{E}\kern-.125emX}}

\IEEEoverridecommandlockouts
    
\input{macros}

\begin{document}
\title{Analyzing and Supporting Adaptation of Online Code Examples}

\author{\IEEEauthorblockN{Tianyi Zhang\IEEEauthorrefmark{2}~~~Di Yang\IEEEauthorrefmark{4}~~~Crista Lopes\IEEEauthorrefmark{4}~~~Miryung Kim\IEEEauthorrefmark{2}}
\thanks{* Both the first author and the second author contributed significantly and this research is led by UCLA.} 
\IEEEauthorblockA{\IEEEauthorrefmark{2}University of California, Los Angeles}~~~\IEEEauthorblockA{\IEEEauthorrefmark{4}University of California, Irvine}
\{tianyi.zhang, miryung\}@cs.ucla.edu~~~\{diy4, lopes\}@uci.edu}

\maketitle

\begin{abstract}
Developers often resort to online Q\&A forums such as Stack Overflow (SO) for filling their programming needs. Although code examples on those forums are good starting points, they are often incomplete and inadequate for developers' local program contexts; adaptation of those examples is necessary to integrate them to production code. 
As a consequence, the process of adapting online code examples is done over and over again, by multiple developers independently. Our work extensively studies these adaptations and variations, serving as the basis for a tool that helps integrate these online code examples in a target context in an interactive manner.

	We perform a large-scale empirical study about the nature and extent of adaptations and variations of SO snippets. We construct a comprehensive dataset linking SO posts to GitHub counterparts based on clone detection, time stamp analysis, and explicit URL references. We then qualitatively inspect 400 SO examples and their GitHub counterparts and develop a taxonomy of 24 adaptation types. Using this taxonomy, we build an automated adaptation analysis technique on top of GumTree to classify the entire dataset into these types. We build a Chrome extension called {\tool} that automatically lifts an adaptation-aware template from each SO example and its  GitHub counterparts to identify hot spots where most changes happen. 
	A user study with sixteen programmers shows that seeing the commonalities and variations in similar GitHub counterparts increases their confidence about the given SO example, and helps them grasp a more comprehensive view about how to reuse the example differently and avoid common pitfalls.

\end{abstract}


\section{Introduction}

Nowadays, a common way of quickly accomplishing programming tasks is to search and reuse code examples in online Q\&A forums such as Stack Overflow (SO)~\cite{umarji2008archetypal, gallardo2009internet, brandt2009two}.  A case study at Google shows that developers issue an average of twelve code search queries per weekday~\cite{sadowski2015developers}. As of July 2018, Stack Overflow has accumulated 26M answers to 16M programming questions. Copying code examples from Stack Overflow is common~\cite{baltes2018usage} and adapting them to fit a target program is recognized as a top barrier when reusing code from Stack Overflow~\cite{wu2018developers}. SO examples are created for illustration purposes, which can serve as a good starting point. However, these examples may be insufficient to be ported to a production environment, as previous studies find that SO examples may suffer from API usage violations~\cite{icse2018}, insecure coding practices~\cite{fischer2017stack}, unchecked obsolete usage~\cite{zhou2016api}, and incomplete code fragments~\cite{treude2017understanding}. Hence, developers may have to manually adapt code examples when importing them into their own projects.

Our goal is to investigate the common adaptation types and their frequencies in online code examples, such as those found in Stack Overflow, which are used by a large number of software developers around the world. To study how they are adopted and adapted in real projects, we contrast them against similar code fragments in GitHub projects. The insights gained from this study could inform the design of tools for helping developers adapt code snippets they find in Q\&A sites. In this paper, we describe one such tool we developed, {\tool}, which works as a Chrome extension.

In broad strokes, the design and main results of our study are as follows. We link SO examples to GitHub counterparts using multiple complementary filters. First, we quality-control GitHub data by removing forked projects and selecting projects with at least five stars. Second, we perform clone detection~\cite{sajnani2016sourcerercc} between 312K SO posts and 51K non-forked GitHub projects to ensure that SO examples are similar to GitHub counterparts. Third, we perform timestamp analysis to ensure that GitHub counterparts are created later than the SO examples. Fourth, we look for explicit URL references from GitHub counterparts to SO examples by matching the post ID. As the result, we construct a comprehensive dataset of {\em variations} and {\em adaptations}. 

When we use all four filters above, we find only 629 SO examples with GitHub counterparts. Recent studies find that very few developers explicitly attribute to the original SO post when reusing code from Stack Overflow~\cite{baltes2018usage, an2017stack, wu2018developers}.
%A recent study finds that, despite the Stack Overflow's licensing terms that require developers to explicitly reference the original question and answer, only 1.8\% repositories containing code from SO follow the licensing policy properly~\cite{baltes2018usage, an2017stack}. Almost one half of the surveyed developers admit copying code from SO without attribution and two thirds are not aware of the SO licensing implications. 
Therefore, we use this resulting set of 629 SO examples as an {\em under-approximation} of SO code reuse and call it an {\em adaptations} dataset. If we apply only the first three filters above, we find 14,124 SO examples with GitHub counterparts that represent potential code reuse from SO to GitHub. While this set does not necessarily imply any causality or intentional code reuse, it still demonstrates the kinds of common variations between SO examples and their GitHub counterparts, which developers might want to consider during code reuse. Therefore, we consider this second dataset as an {\em over-approximation} of SO code reuse, and call it simply a {\em variations} dataset. 

We randomly select 200 clone pairs from each dataset and manually examine the program differences between SO examples and their GitHub counterparts. Based on the manual inspection insights, we construct an adaptation taxonomy with 6 high-level categories and 24 specialized types. We then develop an automated adaptation analysis technique built on top of GumTree~\cite{falleri2014fine} to categorize syntactic program differences into different adaptation types. The precision and recall of this technique are 98\% and 96\% respectively. This technique allows us to quantify the extent of common adaptations and variations in each dataset. The analysis shows that both the adaptations and variations between SO examples and their GitHub counterparts are prevalent and non-trivial. It also highlights several adaptation types such as type conversion, handling potential exceptions, and adding {\ttt if} checks, which are frequently performed yet not automated by existing code integration techniques~\cite{cottrell2008semi, wightman2012snipmatch}. 

Building on this adaptation analysis technique, we develop a Chrome extension called {\tool} to guide developers in adapting and customizing online code examples to their own contexts. For a given SO example, {\tool} shows a list of similar code snippets in GitHub and also lifts an adaptation-aware template from those snippets by identifying common, unchanged code, and also the hot spots where most changes happen. Developers can interact and customize these lifted templates by selecting desired options to fill in the hot spots.
%In the user study, participants using {\tool} annotate 45\% more locations to modify and are able to provide a more comprehensive view of how to use APIs differently or how to avoid common pitfalls. 
We conduct a user study with sixteen developers to investigate whether {\tool} inspires them with new adaptations that they may otherwise ignore during code reuse. Our key finding is that participants using {\tool} focus more on adaptations about code safety (e.g., adding an if check) and logic customization, while participants without {\tool} make more shallow adaptations such as variable renaming. In the post survey, participants find {\tool} help them easily reach consensus on how to reuse a code example, by seeing the commonalities and variations between the example and its GitHub counterparts. Participants also feel more confident after seeing how other GitHub developers use similar code in different contexts, which one participant describes as ``{\em asynchronous pair programming}.''

%In the current
%dataset behind {\tool}, lifted templates contain an average of 35 lines of
%code and 10 hot spots. On average, there are 7 concrete code options for each hot spot.

%We evaluate {\tool} on the dataset of \todo{X} cloned SO examples and find that {\tool} can successfully construct templates for \todo{Y} SO examples. 
%of which \todo{X}\% remain unchanged while \todo{100-X}\% are changed in terms of AST nodes. 

In summary, this work makes the following contributions:
\begin{itemize}
	\item It makes publicly available a comprehensive dataset of {\em adaptations} and {\em variations} between SO and GitHub.\footnote{Our dataset and tool are available at \url{https://github.com/tianyi-zhang/ExampleStack-ICSE-Artifact}} The adaptation dataset includes 629 groups of GitHub counterparts with explicit references to SO posts, and the variation dataset includes 14,124 groups. These datasets are created with care using multiple complementary methods for quality control\textemdash clone detection, time stamp analysis, and explicit references. %A large number of samples are manually inspected by the authors.  
  \item It puts forward an adaptation taxonomy of online code examples and an automated technique for classifying adaptations. This taxonomy is sufficiently different from other change type taxonomies from refactoring~\cite{Fowler2000} and software evolution~\cite{kim2006micro,Fluri2006}, and it captures the particular kinds of adaptations done over online code examples.
  \item It provides browser-based tool support, called {\tool} that displays the commonalities and variations between a SO example and its GitHub counterparts along with their adaptation types and frequencies. Participants find that seeing GitHub counterparts increases their confidence of reusing a SO example and helps them understand different reuse scenarios and corner cases.

\end{itemize}

The rest of the paper is organized as follows. Section~\ref{sec:dataset} describes the data collection pipeline and compares the characteristics of the two datasets. Section~\ref{sec:analysis} describes the adaptation taxonomy development and an automated adaptation analysis technique. Section~\ref{sec:empirical} describes the quantitative analysis of adaptations and variations. Section~\ref{sec:tool} explains the design and implementation of {\tool}. Section~\ref{sec:user-study} describes a user study that evaluates the usefulness of {\tool}. Section~\ref{sec:discussion} discusses threats to validity. Section~\ref{sec:relwork} discusses related work, and Section~\ref{sec:conclusion} concludes the paper.

%\noindent{\em Future Implications.} This work provides empirical evidence about the adaptations between curated examples on Stack Overflow and GitHub projects with potential reuse opportunities (i.e., cloning locations). The Chrome extension, {\tool}, should provide more concrete guidance on how developers should adapt an online example to their own context, by visualizing the commonality and variations between the example and other similar GitHub snippets. In another perspective, {\tool} also has the potential to help the Stack Overflow community to improve the quality of code examples by showing similar GitHub snippets in real-world applications.
%Currently, it is left to a developer to read a curated example and identify which parts are the gist of the example that must be ported versus which parts are simply added for illustration purposes and thus should be removed or modified. {\tool} manifests potential adaptation options in a template with color-coded adaptation types. Furthermore, it allows a user to choose among frequent adaptation options inferred through SO and GitHub cloning analysis.

\section{Linking Stack Overflow to GitHub}
\label{sec:dataset}

This section describes the data collection pipeline. Due to the large portion of unattributed SO examples in GitHub~\cite{baltes2018usage, wu2018developers, an2017stack}, it is challenging to construct a complete set of reused code from SO to GitHub. %Only analyzing GitHub files with SO links will omit a large portion of reused SO code without explicit attribution and skew the result. 
To overcome this limitation, we apply four quality-control filters to {\em underapproximate} and {\em overapproximate} code examples reused from SO to GitHub, resulting in two complementary datasets.

%This section describes the data collection pipeline. Recent studies on the usage and attribution of SO snippets show that very few developers explicitly attribute the original SO post when reusing code from Stack Overflow~\cite{baltes2018usage, wu2018developers, an2017stack}. Therefore, it is hard to estimate a complete set of reused code from SO to GitHub. %Only analyzing GitHub files with SO links will omit a large portion of reused SO code without explicit attribution and skew the result. 
%To overcome this limitation, we decide to use clone detection to find similar code between Stack Overflow and GitHub first, which is also adopted by many other code reuse studies~\cite{baltes2018usage, fischer2017stack, an2017stack, abdalkareem2017code}. We further filter the detected clones to eliminate GitHub clones that are committed before the corresponding SO posts, since they are unlikely to be reused from SO. Finally, we detect a subset of SO examples that are explicitly attributed by GitHub clones, in order to validate the representativeness of the analysis result on potentially reused SO examples detected by clone detection.

%\begin{figure}[!h]
%\centering
%\includegraphics[width=\linewidth]{figures/clone-detection-pipeline-v2.pdf}
%\caption{Data Collection Pipeline}\todo{Update this figure and also show the process of finding clones with SO links.}
%\label{fig:pipeline}
%\end{figure} 

\begin{figure*}[!th]
\begin{minipage}[c]{\linewidth}
  \begin{subfigure}[b]{0.33\linewidth}
    \includegraphics[width=\linewidth]{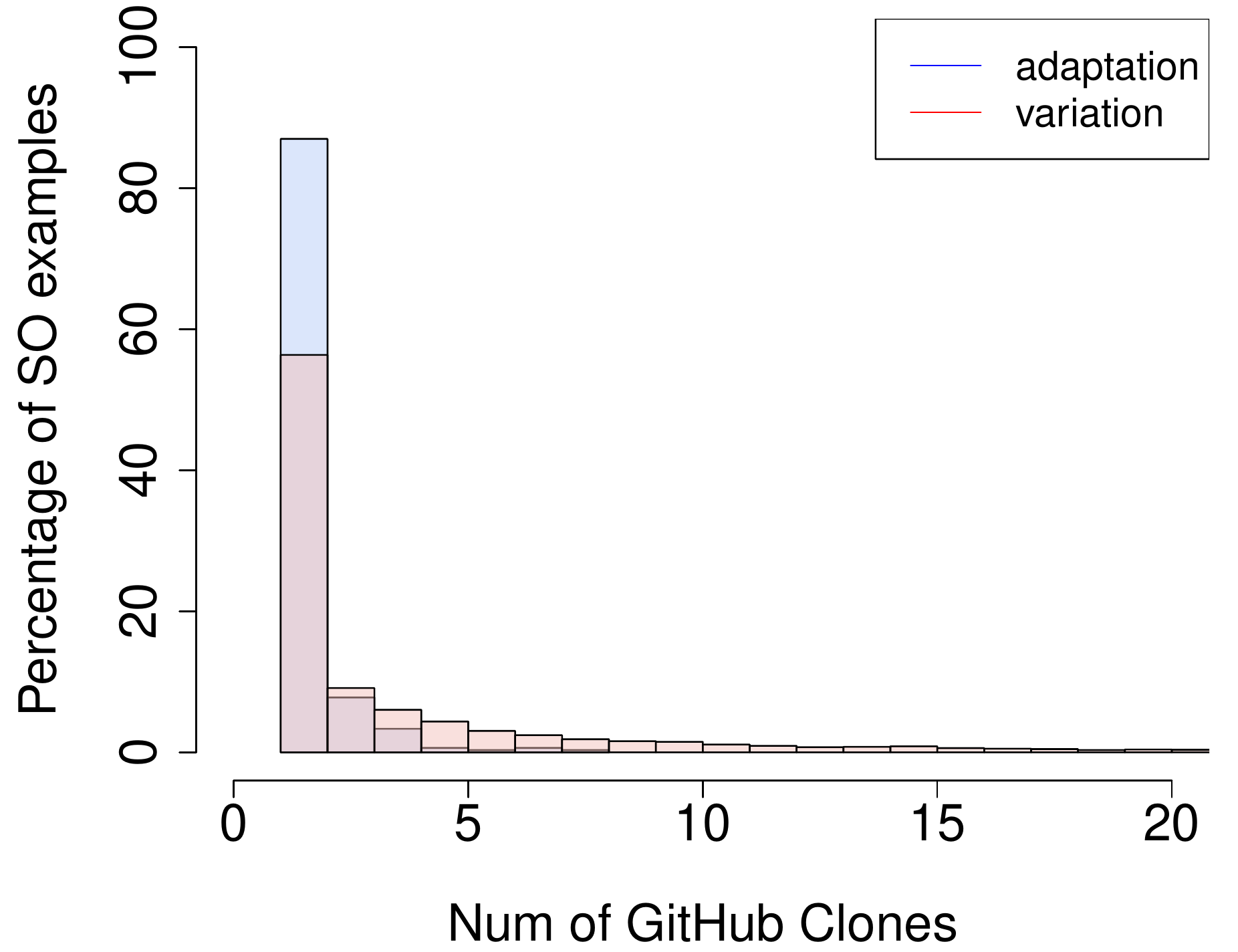}
    \caption{Examples with different numbers of clones}
    \label{fig:clone-dist}
  \end{subfigure}
  \begin{subfigure}[b]{0.33\linewidth}
    \includegraphics[width=\linewidth]{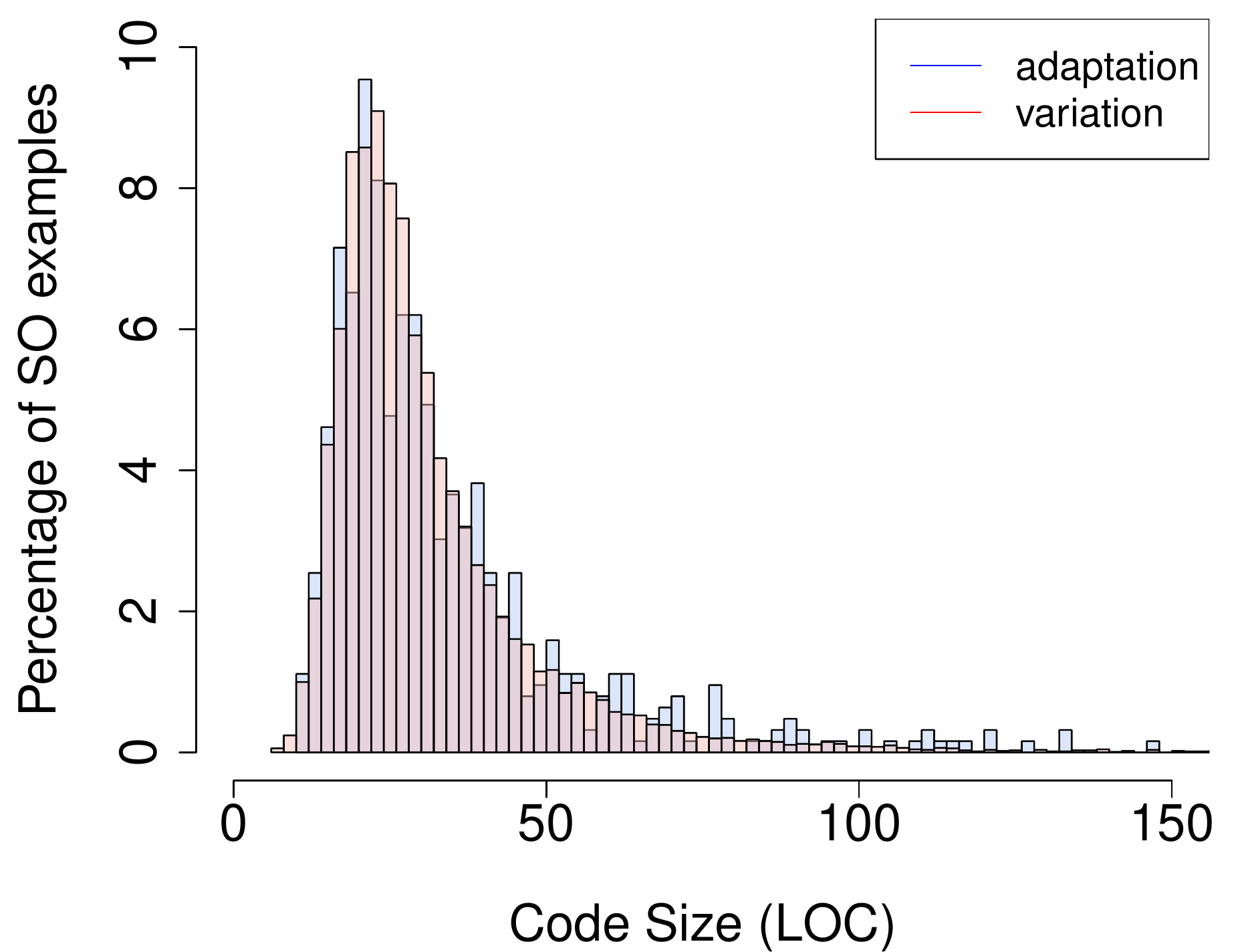}
    \caption{Examples with different code sizes}
    \label{fig:size-dist}
  \end{subfigure}
  \begin{subfigure}[b]{0.33\linewidth}
    \includegraphics[width=\linewidth]{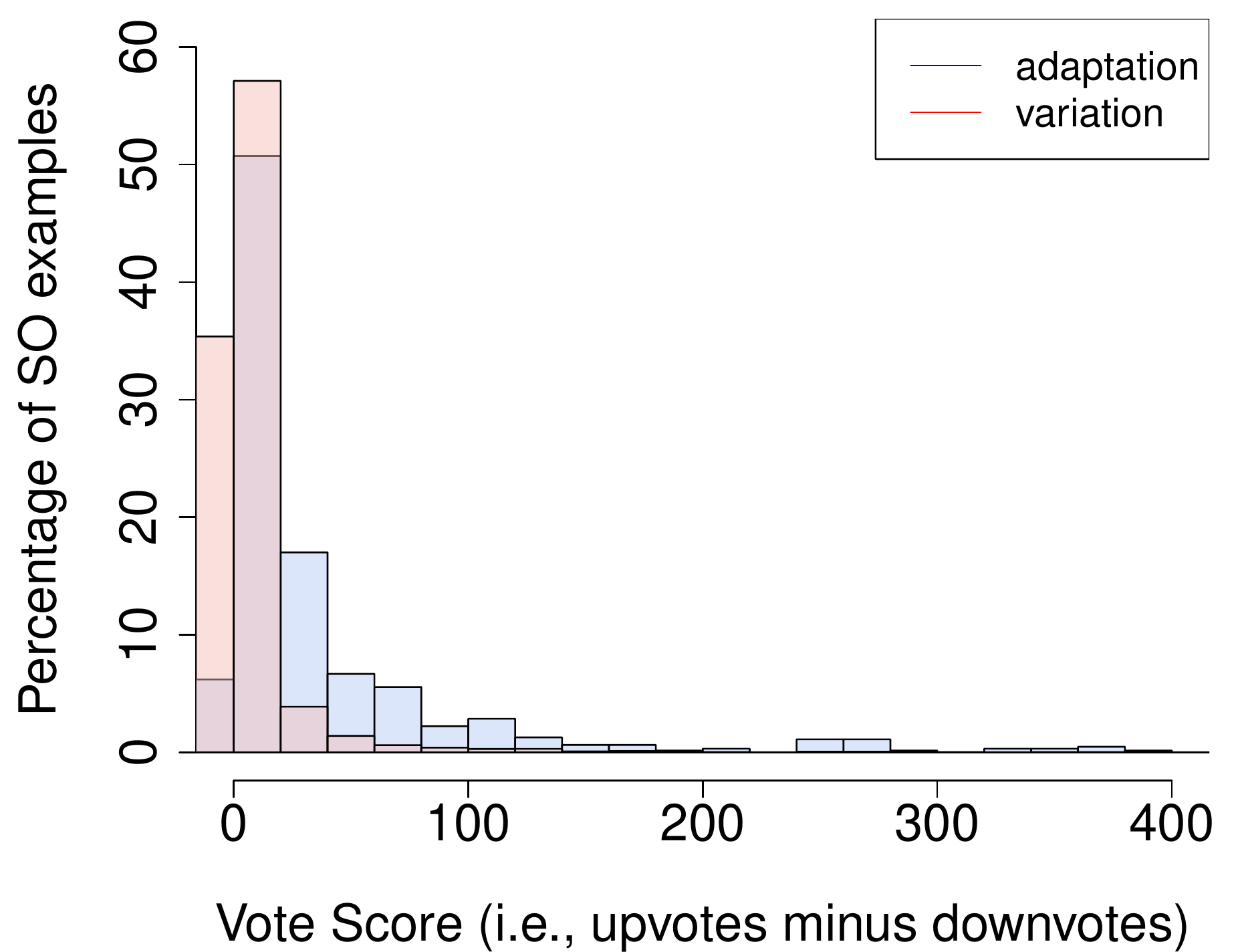}
    \caption{Examples with different vote scores}
    \label{fig:vote-dist}
  \end{subfigure} 

\end{minipage}
\caption{Comparison between SO examples in the adaptation dataset and the variation dataset}
\label{fig:comparison}
\end{figure*}

%\subsection{Clone Detection between SO and GitHub}
%\label{sec:dataset2}
\vspace{1mm}
\noindent{\bf\em GitHub project selection and deduplication.} Since GitHub has many toy projects that do not adequately reflect software engineering practices~\cite{kalliamvakou2014promises}, we only consider GitHub projects that have at least five stars. To account for internal duplication in GitHub~\cite{lopes2017dejavu}, we choose non-fork projects only and further remove duplicated GitHub files using the same file hashing method as in~\cite{lopes2017dejavu}, since such file duplication may skew our analysis. As a result, we download 50,826 non-forked Java repositories with at least five stars from GitTorrent~\cite{gousios2012ghtorrent}. After deduplication, 5,825,727 distinct Java files remain. 

%\noindent\textbf{GitHub snippets.} Our GitHub corpus includes 50,826 non-forked Java repositories with at least five stars, downloaded from GitTorrent~\cite{gousios2012ghtorrent}. We only consider non-fork projects because project forking leads to many identical projects and would unnecessarily skew our analysis. We remove the projects with less than 5 stars, since GitHub has many toy projects that do not adequately reflect software engineering practices~\cite{kalliamvakou2014promises}. Prior work on GitHub cloning finds many identical files among GitHub projects, since developers may copy the whole file into another project without making any changes~\cite{lopes2017dejavu}. After removing identical files, we have 5,825,727 distinct Java files. We extract 9,171,815 Java methods from these distinct files.

%\vspace{1mm}
\noindent{\bf\em Detecting GitHub candidates for SO snippets.} From the SO dump taken in October 2016~\cite{so-data-dump}, we extract 312,219 answer posts that have {\ttt java} or {\ttt android} tags and also contain code snippets in the {\ttt <code>} markdown. We consider code snippets in answer posts only, since snippets in question posts are rarely used as examples. Then we use a token-based clone detector, SourcererCC (SCC)~\cite{sajnani2016sourcerercc} to find similar code between 5.8M distinct Java files and 312K SO posts. We choose SCC because it has high precision and recall and also scales to a large code corpus. Since SO snippets are often free-standing statements~\cite{subramanian2013making,yang2016query}, we parse and tokenize them using a customized Java parser~\cite{subramanian2014live}. Prior work finds that larger SO snippets have more meaningful clones in GitHub~\cite{yang2017stack}. Hence, we choose to study SO examples with no less than 50 tokens, not including code comments, Java keywords, and delimiters. We set the similarity threshold to 70\% since it yields the best precision and recall on multiple clone benchmarks~\cite{sajnani2016sourcerercc}. We cannot set it to 100\% since SCC will then only retain exact copies and exclude those adapted code. We run SCC on a server machine with 116 cores and 256G RAM. It takes 24 hours to complete, resulting in 21,207 SO methods that have one or more similar code fragments (i.e., clones) in GitHub.

\noindent{\bf\em Timestamp analysis.} If the GitHub clone of a SO example is created before the SO post, we consider it unlikely to be reused from SO and remove it from our dataset. To identify the creation date of a GitHub clone, we write a script to retrieve the Git commit history of the file and match the clone snippet against each file revision. We use the timestamp of the earliest matched file revision as the creation time of a GitHub clone. As a result, 7,083 SO examples (33\%) are excluded since all their GitHub clones are committed before the SO posts. %The final dataset includes 14,124 SO examples that are potentially reused to GitHub.

%According to a recent study on the usage and attribution of SO snippets, despite the Stack Overflow's licensing terms that require developers to explicitly reference the original question or answer, 72.5\% of 380 surveyed developers were not aware of SO licensing implications and did not attribute code examples reused from SO~\cite{wu2018developers}. Similarly, in another study, almost one half of surveyed developers admitted copying code from SO without attribution~\cite{baltes2018usage}. Due to the large portion of unattributed code examples, our analysis focuses on potentially reused SO examples detected by SourcererCC and further pruned by checking creation time. 

%\vspace{1mm}
\noindent{\bf\em Scanning explicitly attributed SO examples.} Despite the large portion of unattributed SO examples, it is certainly possible to scan GitHub clones for explicit references such as SO links in code comments to confirm whether a clone is copied from SO. If the SO link in a GitHub clone points to a question post instead of an answer post, we check whether the corresponding SO example is from any of its answer posts by matching the post ID. We find 629 explicitly referenced SO examples. %We randomly sample 200 of them and manually check whether a SO example is correctly matched to a GitHub clone via a SO link. 182 examples are correctly matched, achieving 91\% precision. In the rest 18 examples, though a SO example is in the same discussion thread or the same answer post referenced by a GitHub file, we find another similar example in the same thread or answer post that looks more like the origin of the GitHub clone.
 
%By scanning the GitHub files in our corpus, we find \todo{X} files that reference Stack Overflow via a link in code comments. Note that a SO link may point to an entire discussion thread with multiple SO posts, instead of a specific post where the code is copied from. In another case, a SO post may contain multiple code snippets, while a GitHub file may only copy one snippet. Therefore, we need to further perform code matching to identify which specific code snippets are copied. We use the Jaccard distance to measure the code similarity between a SO snippet and a GitHub file. More specifically, if a SO link appears before a class, we find the best match for each method in the GitHub file. If a SO link appears before a method, we consider matching that single method only. If a SO link appears in the middle of a method, we consider matching code after the SO link only. If a SO code snippet is a list of free-standing program statements without a method header, we consider it as a single method. As a result, we find \todo{X} matched method pairs between the GitHub files and their SO references. We manually inspect a sample of \todo{Y} GitHub files and their references and find that our code matching strategy can accurately identify the copied code snippets with \todo{A}\% precision and \todo{B}\% recall.

%\vspace{1mm}
\noindent{\bf\em Overapproximating and underapproximating reused code.} We use the set of 629 explicitly attributed SO examples as an {\em underapproximation} of reused code from SO to GitHub, which we call an {\em adaptation} dataset. 
We consider the 14,124 SO examples after timestamp analysis as an {\em overapproximation} of potentially reused code, which we call a {\em variation} dataset. Figure~\ref{fig:comparison} compares the characteristics of these two datasets of SO examples in terms of the number of GitHub clones, code size, and vote score (i.e., upvotes minus downvotes). Since developers do not often attribute SO code examples, explicitly referenced SO examples have a median of one GitHub clone only, while SO examples have a median of two clones in the variation dataset. Both sets of SO examples have similar length, 26 vs.~25 lines of code in median. However, SO examples from the adaptation dataset have significantly more upvotes than the variation dataset: 16 vs.~1 in median. In the following sections, we inspect, analyze, and quantify the adaptations and variations evidenced by both datasets.

% Please add the following required packages to your document preamble:
% \usepackage{multirow}
\begin{table*}%[!tbh]
\centering
\caption{Common adaptation types, categorization, and implementation}
\label{tab:types}
\resizebox{\textwidth}{!}{%
\renewcommand{\arraystretch}{1.3}
\renewcommand\tabcolsep{4pt}
\begin{tabular}{|l|l|l|}
\hline
\multicolumn{1}{|c|}{\bf Category}              & \multicolumn{1}{c|}{\bf Adaptation Type}                       & \multicolumn{1}{c|}{\bf Rule}                                                                                                                                                   \\ \hline
\multirow{4}{*}{Code Hardening}                                                        & Add a conditional                                          & Insert($t_1$, $t_2$, $i$) $\wedge$ NodeType($t_1$, IfStatement)                                                                                                                                                                                          \\ \cline{2-3} 
                                                                                       & Insert a final modifier                                    & Insert($t_1$, $t_2$, $i$) $\wedge$ NodeType($t_1$, Modifier) $\wedge$ NodeValue($t_1$, final)                                                                                                                                                             \\ \cline{2-3} 
                                                                                       & Handle a new exception type                                & Exception(e, GH) $\wedge$ $\neg$Exception(e, SO)                                                                                                                                                                        \\ \cline{2-3} 
                                                                                       & Clean up unmanaged resources (e.g. close a stream)         & (LocalCall(m, GH) $\vee$ InstanceCall(m, GH)) $\wedge$ $\neg$LocalCall(m, SO) $\wedge$ $\neg$InstanceCall(m, SO) $\wedge$ isCleanMethod(m)                                                                     \\ \hline
\multirow{3}{*}{\begin{tabular}[c]{@{}l@{}}Resolve Compilation \\ Errors\end{tabular}} & Declare an undeclared variable                             & Insert($t_1$, $t_2$, $i$) $\wedge$ NodeType($t_1$, VariableDeclaration) $\wedge$ NodeValue($t_1$, v) $\wedge$ Use(v, SO) $\wedge$ $\neg$Def(v, SO)                                                                                                 \\ \cline{2-3} 
                                                                                       & Specify a target of method invocation                      & InstanceCall(m, GH) $\wedge$ LocalCall(m, SO)                                                                                                                                                                     \\ \cline{2-3} 
                                                                                       & Remove undeclared variables or local method calls          & (Use(v, SO) $\wedge$ $\neg$Def(v, SO) $\wedge$ $\neg$Use(v, GH)) $\vee$
 (LocalCall(m, SO) $\wedge$ $\neg$LocalCall(m, GH) $\wedge$ $\neg$InstanceCall(m, GH))                                          \\ \hline
\multirow{5}{*}{Exception Handling}                                                    & Insert/delete a try-catch block                            & (Insert($t_1$, $t_2$, $i$) $\vee$ Delete($t_1$)) $\wedge$ NodeType($t_1$, TryStatement)                                                                                                                                                                     \\ \cline{2-3} 
                                                                                       & Insert/delete a thrown exception in a method header        & \begin{tabular}[c]{@{}l@{}}Changed($t_1$) $\wedge$ NodeType($t_1$, Type) $\wedge$ Parent($t_2$, $t_1$) $\wedge$ NodeType($t_2$, MethodDeclaration) $\wedge$ NodeValue($t_1$, t) $\wedge$ \\ isExceptionType(t)\end{tabular}                                     \\ \cline{2-3} 
                                                                                       & Update the exception type                                  & \begin{tabular}[c]{@{}l@{}}Update($t_1$, $t_2$) $\wedge$ NodeType($t_1$, SimpleType) $\wedge$ NodeType($t_2$, SimpleType) $\wedge$ NodeValue($t_1$, $v_1$) $\wedge$ \\ isExceptionType($v_1$) $\wedge$ NodeValue($t_2$, $v_2$) $\wedge$ isExceptionType($v_2$)\end{tabular} \\ \cline{2-3} 
                                                                                       & Change statements in a catch block                         & Changed($t_1$) $\wedge$ Ancestor($t_2$, $t_1$) $\wedge$ NodeType($t_2$, CatchClause)                                                                                                                                                                      \\ \cline{2-3} 
                                                                                       & Change statements in a finally block                       & Changed($t_1$) $\wedge$ Ancestor($t_2$, $t_1$) $\wedge$ NodeType($t_2$, FinallyBlock)                                                                                                                                                                     \\ \hline
\multirow{4}{*}{Logic Customization}                                                   & Change a method call                                       & Changed($t_1$) $\wedge$ Ancestor($t_2$, $t_1$) $\wedge$ NodeType($t_2$, MethodInvocation)                                                                                                                                                                 \\ \cline{2-3} 
                                                                                       & Update a constant value                                    & Update($t_1$, $t_2$) $\wedge$ NodeType($t_1$, Literal) $\wedge$ NodeType($t_2$, Literal)                                                                                                                                                                  \\ \cline{2-3} 
                                                                                       & Change a conditional expression                            & \begin{tabular}[c]{@{}l@{}}Changed($t_1$) $\wedge$  Ancestor($t_2$, $t_1$) $\wedge$\\ (NodeType($t_2$, IfCondition) $\vee$ NodeType($t_2$, LoopCondition) $\vee$ NodeType($t_2$, SwitchCase))\end{tabular}                                                                                                \\ \cline{2-3} 
                                                                                       & Change the type of a variable                              & Update($t_1$, $t_2$) $\wedge$ NodeType($t_1$, Type) $\wedge$ NodeType($t_2$, Type)                                                                                                                                                                        \\ \hline
\multirow{3}{*}{Refactoring}                                                           & Rename a variable/field/method                             & Update($t_1$, $t_2$) $\wedge$ NodeType($t_1$, Name)                                                                                                                                                                                                    \\ \cline{2-3} 
                                                                                       & Replace hardcoded constant values with variables           & Delete($t_1$) $\wedge$ NodeType($t_1$, Literal) $\wedge$Insert($t_1$, $t_2$, $i$) $\wedge$ NodeType($t_1$, Name) $\wedge$ Match($t_1$, $t_2$)                                                                                                                        \\ \cline{2-3} 
                                                                                       & Inline a field                                             & Delete($t_1$) $\wedge$ NodeType($t_1$, Name) $\wedge$Insert($t_1$, $t_2$, $i$) $\wedge$ NodeType($t_1$, Literal) $\wedge$ Match($t_1$, $t_2$)                                                                                                                        \\ \hline
\multirow{4}{*}{Miscellaneous}                                                         & Change access modifiers                                    & Changed($t_1$) $\wedge$ NodeType($t_1$, Modifier) $\wedge$ NodeValue($t_1$, v) $\wedge$ v $\in$ \{private, public, protected, static\}                                                                                                                 \\ \cline{2-3} 
                                                                                       & Change a log/print statement                               & Changed($t_1$) $\wedge$ NodeType($t_1$, MethodInvocation) $\wedge$ NodeValue($t_1$, m) $\wedge$ isLogMethod(m)                                                                                                                                         \\ \cline{2-3} 
                                                                                       & Style reformatting (i.e., inserting/deleting curly braces) & Changed($t_1$) $\wedge$ NodeType($t_1$, Block) $\wedge$ Parent($t_2$, $t_1$) $\wedge$ $\neg$Changed($t_2$) $\wedge$ Child($t_3$, $t_1$) $\wedge$ $\neg$Changed($t_3$)                                                                                                 \\ \cline{2-3} 
                                                                                       & Change Java annotations                                    & Changed($t_1$) $\wedge$ NodeType($t_1$, Annotation)  \\ \cline{2-3} 
                                                                                                                                                                                                                                                                                                                                                                              & Change code comments                                   & Changed($t_1$) $\wedge$ NodeType($t_1$, Comment) 
 \\ \hline
\end{tabular}
}%
\newline
\vspace{2mm}
\resizebox{\textwidth}{!}{%
\renewcommand{\arraystretch}{1.3}
\renewcommand\tabcolsep{4pt}
\begin{tabular}{|l|l|l|}
\hline
\multicolumn{1}{|c|}{\bf GumTree Edit Operation}                                                                                                                                                                        & \multicolumn{1}{c|}{\bf Syntactic Predicate}                                                                                                                                                                      & \multicolumn{1}{c|}{\bf Semantic Predicate}                                                                                                                   \\ \hline
\multirow{2}{*}{\begin{tabular}[c]{@{}l@{}}{\bf Insert($t_1$, $t_2$, $i$)} inserts a new tree node $t_1$ as the $i$-th \\ child of $t_2$ in the AST of a GitHub snippet.\end{tabular}}                                  & {\bf NodeType($t_1$, $X$)} checks if the node type of $t_1$ is $X$.                                                                                                                                                                 & \begin{tabular}[c]{@{}l@{}}{\bf Exception($e$, $P$)} checks if $e$ is an exception caught in a {\ttt catch}\\ clause or thrown in a method header in program $P$.\end{tabular}     \\ \cline{2-3} 
                                                                                                                                                                                                                    & \begin{tabular}[c]{@{}l@{}}{\bf NodeValue($t_1$, $v$)} checks if the corresponding source code\\ of node $t_1$ is $v$.\end{tabular}                                                                                                & {\bf LocalCall($m$, $P$)} checks if $m$ is a local method call in program $P$.                                                                                               \\ \hline
\multirow{2}{*}{\begin{tabular}[c]{@{}l@{}}{\bf Delete($t$)} removes the tree node $t$ from the AST of a\\ SO example.\end{tabular}}                                                                                                                              & \begin{tabular}[c]{@{}l@{}}{\bf Match($t_1$, $t_2$)} checks if $t_1$ and $t_2$ are matched based on \\ surrounding nodes regardless of node types.\end{tabular}                                                                      & {\bf InstanceCall($m$, $P$)} checks if $m$ is an instance call in program $P$.                                                                                                \\ \cline{2-3} 
                                                                                                                                                                                                                    & {\bf Parent($t_1$, $t_2$)} checks if $t_1$ is the parent of $t_2$ in the AST.                                                                                                                                                         & {\bf Def($v$, $P$)} checks if variable $v$ is defined in program $P$.                                                                                                        \\ \hline
\multirow{2}{*}{\begin{tabular}[c]{@{}l@{}}{\bf Update($t_1$, $t_2$)} updates the tree node $t_1$ in a SO\\ example with $t_2$ in the GitHub counterpart.\end{tabular}}                                      & {\bf Ancestor($t_1$, $t_2$)} checks if $t_1$ is the ancestor of $t_2$ in the AST.                                                                                                                                                     & {\bf Use($v$, $P$)} checks if variable $v$ is used in program $P$.                                                                                                           \\ \cline{2-3} 
                                                                                                                                                                                                                    & {\bf Child($t_1$, $t_2$)} checks if $t_1$ is the child of $t_2$.                                                                                                                                                                      & {\bf IsExceptionType($X$)} checks if $X$ contains ``Exception''.                                                                                               \\ \hline
\multirow{2}{*}{\begin{tabular}[c]{@{}l@{}}{\bf Move($t_1$, $t_2$, $i$)} moves an existing node $t_1$ in the\\ AST of a SO example as the $i$-th child of $t_2$ in \\ the GitHub counterpart.\end{tabular}} & \multirow{2}{*}{\begin{tabular}[c]{@{}l@{}}{\bf Changed($t_1$)} is a shorthand for {\bf Insert($t_1$, $t_2$, $i$)} $\vee$ {\bf Delete($t_1$)} \\ $\vee$ {\bf Update($t_1$, $t_2$)} $\vee$ {\bf Move($t_1$, $t_2$)}, which checks any \\ edit operation on $t_1$.\end{tabular}} & \begin{tabular}[c]{@{}l@{}}{\bf IsLogMethod($X$)} checks if $X$ is one of the predefined log methods, \\ e.g., log, println, error, etc.\end{tabular}                            \\ \cline{3-3} 
                                                                                                                                                                                                                    &                                                                                                                                                                                                                    & \begin{tabular}[c]{@{}l@{}}{\bf IsCleanMethod($X$)} checks if $X$ is one of the predefined resource\\  clean-up methods, e.g., close, recycle, dispose, etc.\end{tabular} \\ \hline
\end{tabular}
}%
\end{table*}

\section{Adaptation Type Analysis}
\label{sec:analysis}
%Section~\ref{sec:inspect} describes manual inspection to produce a taxonomy of adaptation categories and Section~\ref{sec:categorization} describes an automated adaptation categorization technique.  

%%% Section 3.1 Manual Inspection %%%
\subsection{Manual Inspection}
\label{sec:inspect}

To get insights into adaptations and variations of SO examples, we randomly sample SO examples and their GitHub counterparts from each dataset and inspect their program differences using GumTree~\cite{falleri2014fine}. Below, we use ``adaptations'' to refer both adaptations and variations for simplicity.

The first and the last authors jointly labeled these SO examples with adaptation descriptions and grouped the edits with similar descriptions to identify common adaptation types. We initially inspected 90 samples from each dataset and had already observed convergent adaptation types. We continued to inspect more and stopped after inspecting 200 samples from each dataset, since the list of adaptation types was converging. This is a typical procedure in qualitative analysis~\cite{berg2004qualitative}. The two authors then discussed with the other authors to refine the adaptation types. Finally, we built a taxonomy of 24 adaptation types in 6 high-level categories, as shown in Table~\ref{tab:types}.

\noindent{\bf Code Hardening.} This category includes four adaptation types that strengthen SO examples in a target project. {\em Insert a conditional} adds an {\ttt if} statement that checks for corner cases or protects code from invalid input data such as {\ttt null} or an out-of-bound index. {\em Insert a final modifier} enforces that a variable is only initialized once and the assigned value or reference is never changed, which is generally recommended for clear design and better performance due to static inlining. {\em Handle a new exception} improves the reliability of a code example by handling any missing exceptions, since exception handling is often omitted in examples in SO~\cite{icse2018}. {\em Clean up unmanaged resources} helps release unneeded resources such as file streams and web sockets to avoid resource leaks~\cite{torlak2010effective}.

\noindent{\bf Resolve Compilation Errors.} SO examples are often incomplete with undefined variables and method calls~\cite{dagenais2012recovering,yang2016query}. {\em Declare an undeclared variable} inserts a statement to declare an unknown variable. {\em Specify a target of method invocation} resolves an undefined method call by specifying the receiver object of that call. In an example about getting CPU usage~\cite{so-example-cpu}, one comment complains the example does not compile due to an unknown method call, {\ttt getOperatingSystemMXBean}. Another suggests to preface the method call with an instance, {\ttt ManagementFactory}, which is also evidenced by its GitHub counterpart~\cite{gh-clone-cpu}. Sometimes, statements that use undefined variables and method calls are simply deleted.

\noindent{\bf Exception Handling.} This category represents changes of the exception handling logic in {\ttt catch/finally} blocks and  {\ttt throws} clauses. One common change is to customize the actions in a {\ttt catch} block, e.g., printing a short error message instead of the entire stack trace. Some developers handle exceptions locally rather than throwing them in method headers. For example, while the SO example~\cite{so-example-exception} throws a generic {\ttt Exception} in the {\ttt addLibraryPath} method, its GitHub clone~\cite{gh-clone-exception} enumerates all possible exceptions such as {\ttt SecurityException} and {\ttt IllegalArgumentException} in a {\ttt try-catch} block. By contrast, propagating the exceptions to upstream by adding {\tt throws} in the method header is another way to handle the exceptions. 

\noindent{\bf Logic Customization.} Customizing the functionality of a code example to fit a target project is a common and broad category. We categorize logic changes to four basic types. {\em Change a method call} includes any edits in a method call, e.g., adding or removing a method call, changing its arguments or receiver, etc. {\em Update a constant value} changes a constant value such as the thread sleep time to another value. {\em Change a conditional expression} includes any edits on the condition expression of an {\ttt if} statement, a {\ttt loop}, or a {\ttt switch case}. 

{\em Update a type name} replaces a variable type or a method return type with another type. For example, {\ttt String} and {\ttt StringBuffer} appear in multiple SO examples, and a faster type, {\ttt StringBuilder}, is used in their GitHub clones instead. Such type replacement often involves extra changes such as updating method calls to fit the replaced type or adding method calls to convert one type to another. For example, instead of returning {\ttt InetAddress} in a SO example~\cite{so-example-address}, its GitHub clone~\cite{gh-clone-address} returns {\ttt String} and thus converts the IP address object to its string format using a new {\ttt Formatter} API. %Such type conversion can be automated via type analysis.

%%% Move the quantification of adaptation edits here for balanced page layout
\begin{figure*}[!t]
\begin{minipage}[c]{\linewidth}
  \begin{subfigure}[b]{0.33\linewidth}
    \includegraphics[width=\linewidth]{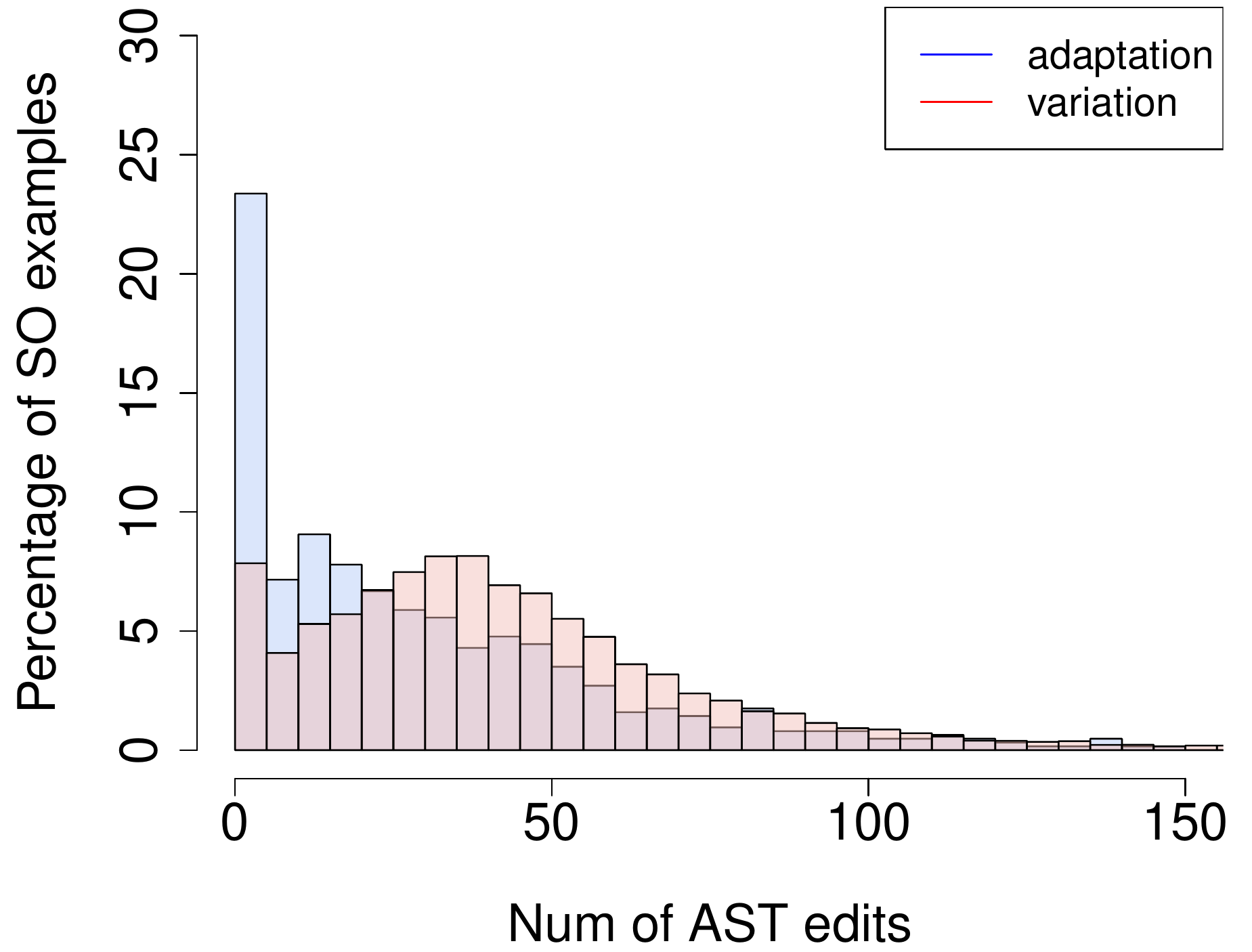}
    \caption{Distribution of AST edits}
    \label{fig:edit-dist}
  \end{subfigure}
  \begin{subfigure}[b]{0.33\linewidth}
    \includegraphics[width=\linewidth]{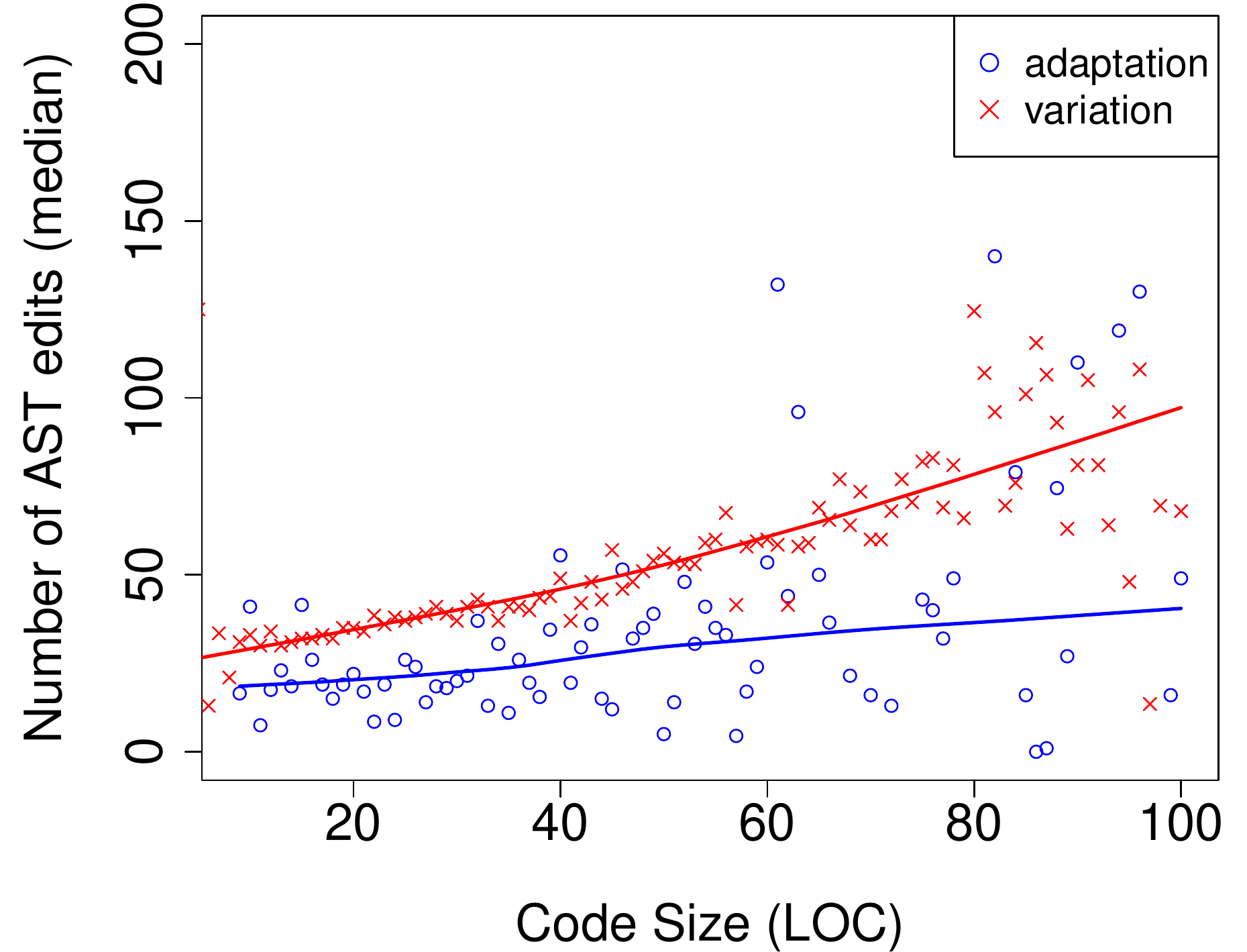}
    \caption{Code size vs.~AST edits}
    \label{fig:size-edit}
  \end{subfigure}
  \begin{subfigure}[b]{0.33\linewidth}
    \includegraphics[width=\linewidth]{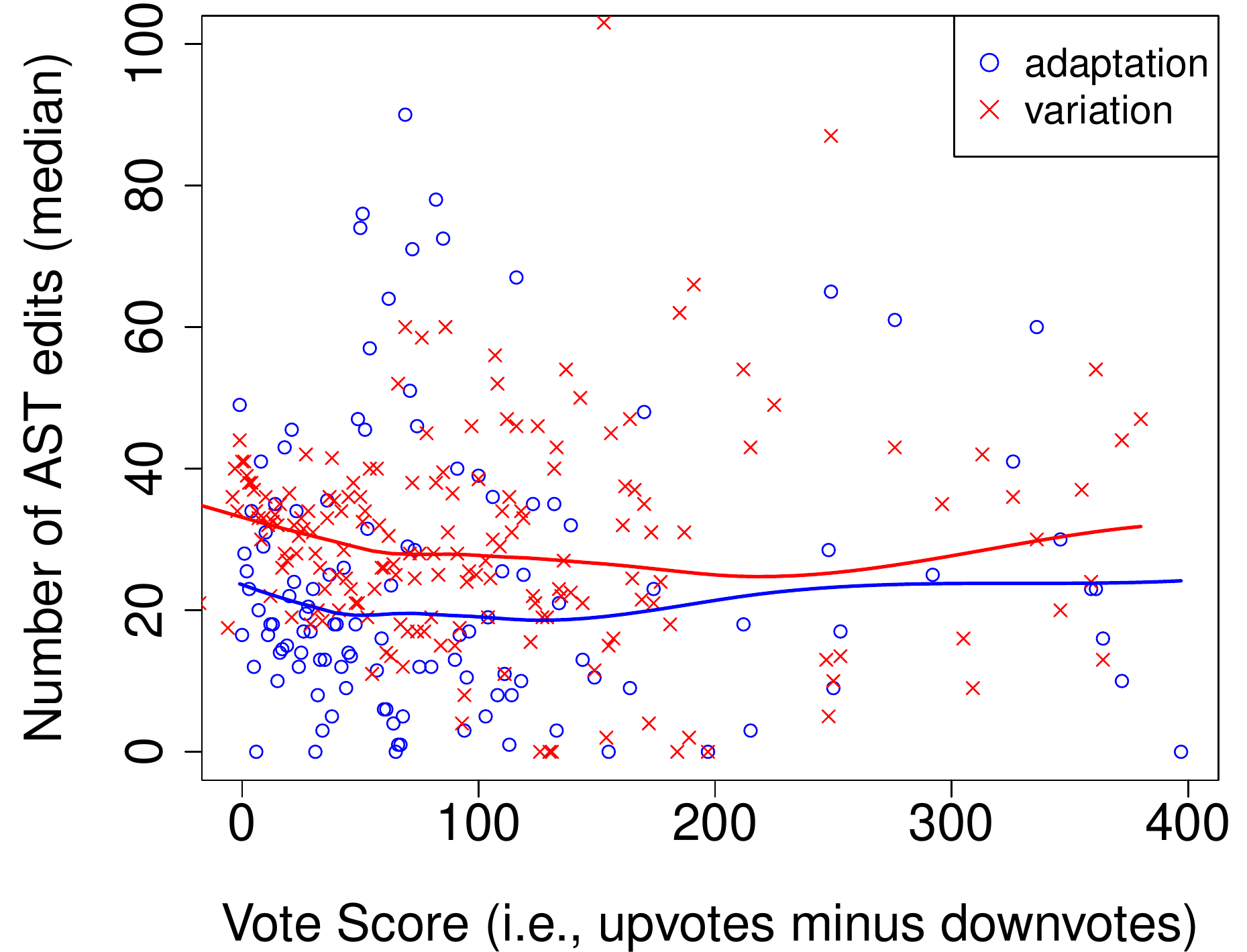}
    \caption{Vote score vs.~AST edits}
    \label{fig:vote-edit}
  \end{subfigure} 
  \caption{Code size (LOC) and vote scores on the number of AST edits in a SO example}
  \end{minipage}
\end{figure*}

\noindent{\bf Refactoring.} 31\% of inspected GitHub counterparts use a method or variable name different from the SO example. Instead of {\ttt slider} in a SO example~\cite{so-example-jslide}, {\ttt timeSlider} is used in one GitHub counterpart~\cite{gh-clone-jslide} and {\ttt volumnSlider} is used in another counterpart~\cite{gh-clone-jslide2}. Because SO examples often use hardcoded constant values for illustration purposes, GitHub counterparts may use variables instead of hardcoded constants. However, sometimes, a GitHub counterpart such as~\cite{gh-clone-download} does the opposite by inlining the values of two constant fields, {\ttt BUFFER_SIZE} and {\ttt KB}, since these fields do not appear along with the copied method, {\ttt downloadWithHttpClient}~\cite{so-example-download}.

\noindent{\bf Miscellaneous.} Adaptation types in this category do not have a significant impact on the reliability and functionality of a SO example. However, several interesting cases are still worth noting. In 91 inspected examples, GitHub counterparts include comments to explain the reused code. Sometimes, annotations such as {\ttt @NotNull} or {\ttt @DroidSafe} appear in GitHub counterparts to document the constraints of code.
%Five participants in our user study also did the same, expressing the desire of not only reusing the functionality, but also being able to explain the reused code. 

%%% Section 3.2 Adaptation Analysis Technique %%%
\subsection{Automated Adaptation Categorization}
\label{sec:categorization}

Based on the manual inspection, we build a rule-based classification technique that automatically categorizes AST edit operations generated by GumTree to different adaptation types. %GumTree computes a set of AST edit operations to transform one program to another. GumTree first traverses the ASTs, finds isomorphic subtrees between ASTs, and builds mappings between tree nodes in each pair of isomorphic subtrees. Then GumTree generates a set of edit operations using a linear optimal algorithm~\cite{Chawathe1996}, based on the established mappings. 
GumTree supports four edit operations---{\bf insert}, {\bf delete}, {\bf update}, and {\bf move}, described in Column {\sf GumTree Edit Operation} in Table~\ref{tab:types}.  Given a set of AST edits, our technique leverages both syntactic and semantic rules to categorize the edits to 24 adaptation types. Column {\sf Rule} in Table~\ref{tab:types} describes the implementation logic of categorizing each adaptation type.

\noindent{\bf Syntactic-based Rules.} 16 adaptation types are detected based on syntactic information, e.g., edit operation types, AST node types and values, etc. Column {\sf Syntactic Predicate} defines such syntactic information, which is obtained using the built-in functions provided by GumTree. For example, the rule {\em insert a final modifier} checks for an edit operation that inserts a {\ttt Modifier} node whose value is {\ttt final} in a GitHub clone. 

\noindent{\bf Semantic-based Rules.} 8 adaptation types require leveraging semantic information to be detected (Column {\sf Semantic Predicate}). For example, the rule {\em declare an undeclared variable} checks for an edit operation that inserts a {\ttt VariableDeclaration} node in the GitHub counterpart and the variable name is {\em used} but not {\em defined} in the SO example. Our technique traverses ASTs to gather such semantic information. For example, our AST visitor keeps track of all declared variables when visiting a {\ttt VariableDeclaration} AST node, and all used variables when visiting a {\ttt Name} node.

%%% Section 3.3 evaluation on the analysis technique
\subsection{Accuracy of Adaptation Categorization}
\label{sec:eval1}

We randomly sampled another 100 SO examples and their GitHub clones to evaluate our automated categorization technique. To reduce bias, the second author who was not involved in the previous manual inspection labeled the adaptation types in this validation set. The ground truth contains 449 manually labeled adaptation types in 100 examples. Overall, our technique infers 440 adaptation types with 98\% precision and 96\% recall. In 80\% of SO examples, our technique infers all adaptation types correctly. In another 20\% of SO examples, it infers some but not all expected adaptation types. 

Our technique infers incorrect or missing adaptation types for two main reasons. First, our technique only considers 24 common adaptation types in Table~\ref{tab:types} but does not handle infrequent ones such as refactoring using lambda expressions and rewriting {\ttt ++i} to {\ttt i++}. Second, GumTree may generate sub-optimal edit scripts with unnecessary edit operations in about 5\% of file pairs, according to \cite{falleri2014fine}. %For example, a GitHub clone inserts an additional clause in the existing {\ttt if} condition and changes it from {\ttt v.contains(...)} to {\ttt x!=null \&\& v.contains(...)}.\footnote{The variable names in the {\tttt if} conditions are simplified for illustrating purposes. The clone pair is the {\tttt setCamera} methods in \url{https://stackoverflow.com/questions/8003222} and in \url{https://github.com/zhenSearcher/native-camera/blob/master/app/src/main/java/com/left/to/zlu/nativecamera/CameraPreview.java\#L66-L82}.} GumTree mistakenly detects the movement of {\ttt contains} as two separate insert and delete operations of the {\ttt contains} method call. As a consequence, our technique mistakenly reports the {\em change a method call} adaptation on {\ttt contains}. 
In such cases, our technique may mistakenly report incorrect adaptation types.

\begin{figure*}[!t]
\begin{minipage}[c]{\linewidth}
  \centering
  \begin{subfigure}[b]{0.49\linewidth}
	\includegraphics[width=\linewidth]{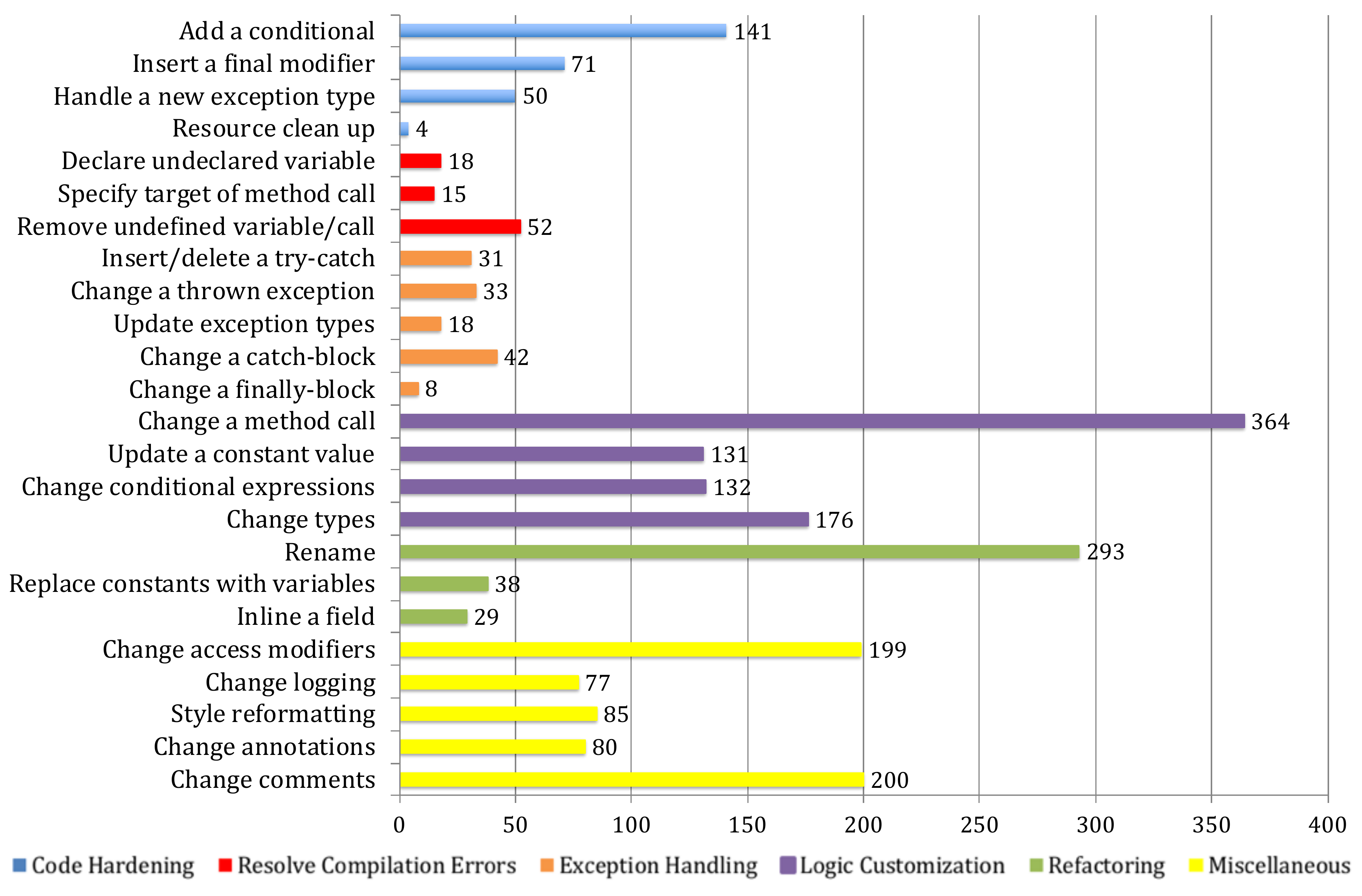}
    \caption{Adaptations: 629 explicitly attributed SO examples}
    \label{fig:type-count-explicit}
  \end{subfigure}
  \begin{subfigure}[b]{0.49\linewidth}
    \includegraphics[width=\linewidth]{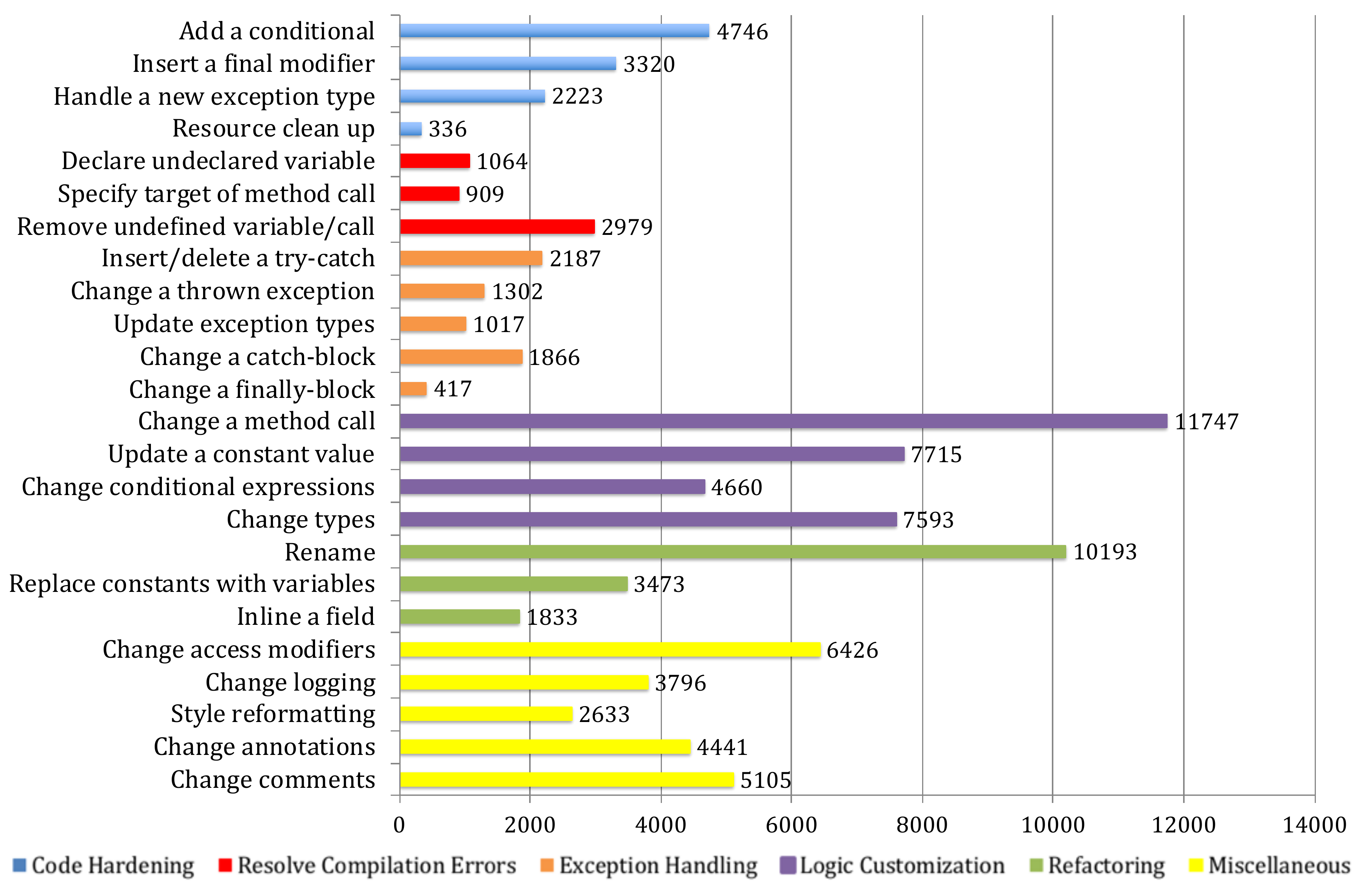}
    \caption{Variations: 14,124 potentially reused SO examples}
    \label{fig:type-count-potential}
   \end{subfigure}
\end{minipage}
\caption{Frequencies of categorized adaptation types in two datasets}
\label{fig:adapt-type-frequency}
\end{figure*} 

\section{Empirical Study}
\label{sec:empirical}

\subsection{How many edits are potentially required to adapt a SO example?}
\label{sec:RQ1}
We apply the adaptation categorization technique to quantify the extent of adaptions and variations in the two datasets. We measure AST edits between a SO example and its GitHub counterpart. If a SO code example has multiple GitHub counterparts, we use the average number. Overall, 13,595 SO examples (96\%) in the variation dataset include a median of 39 AST edits (mean 47). 556 SO examples (88\%) in the adaptation dataset include a median of 23 AST edits (mean 33). Figure~\ref{fig:edit-dist} compares the distribution of AST edits in these two datasets. In both datasets, most SO examples have variations from their counterparts, indicating that integrating them to production code may require some type of adaptations. 

Figure~\ref{fig:size-edit} shows the median number of AST edits in SO examples with different lines of code. We perform a non-parametric local regression~\cite{shyu2017local} on the example size and the number of AST edits. As shown by the two lines in Figure~\ref{fig:size-edit}, there is a strong positive correlation between the number of AST edits and the SO example size in both datasets\textemdash long SO examples have more adaptations than short examples. 
% MK :remove this. you have no evidence to align this finding. It's far too stretched. The difficulty of adapting long code examples is further confirmed by the following user study, where participants state that it is hard to quickly spot where to modify in a lengthy code example (Section~\ref{sec:user}).

Stack Overflow users can vote a post to indicate the applicability and usefulness of the post. Therefore, votes are often considered as the main quality metric of SO examples~\cite{nasehi2012makes}. Figure~\ref{fig:vote-edit} shows the median number of AST edits in SO examples with different vote scores. Although the adaptation dataset has significantly higher votes than the variation dataset (Figure~\ref{fig:vote-dist}), there is no strong positive or negative correlation between the AST edit and the vote score in both sets. This implies that highly voted SO examples do not necessarily require fewer adaptations than those with low vote scores. 

\subsection{What are common adaptation and variation types?}
%\subsection{Quantitative Analysis of Adaptation Categorization}
\label{sec:RQ2}

%To investigate the common adaptation types across different SO examples, we count the distinct types of adaptations applied to a code example. 
Figure~\ref{fig:adapt-type-frequency} compares the frequencies of the 24 categorized adaptation types (Column {\sf Adaptation Type} in Table~\ref{tab:types}) for the adaptation and variation datasets. If a SO code example has multiple GitHub counterparts, we only consider the distinct types among all GitHub counterparts to avoid the inflation caused by repetitive variations among different counterparts. The frequency distribution is consistent in most adaptation types between the two datasets, indicating that {\em variation patterns resemble adaptation patterns}. Participants in the user study (Section~\ref{sec:user-study}) also appreciate being able to see variations in similar GitHub code, since ``it highlights the best practices followed by the community and prioritizes the changes that I should make first,'' as P5 explained.

%%%%% Move the big screenshot here
\begin{figure*}[!t]
\centering    
\resizebox{\textwidth}{!}{\includegraphics[width=0.7\linewidth]{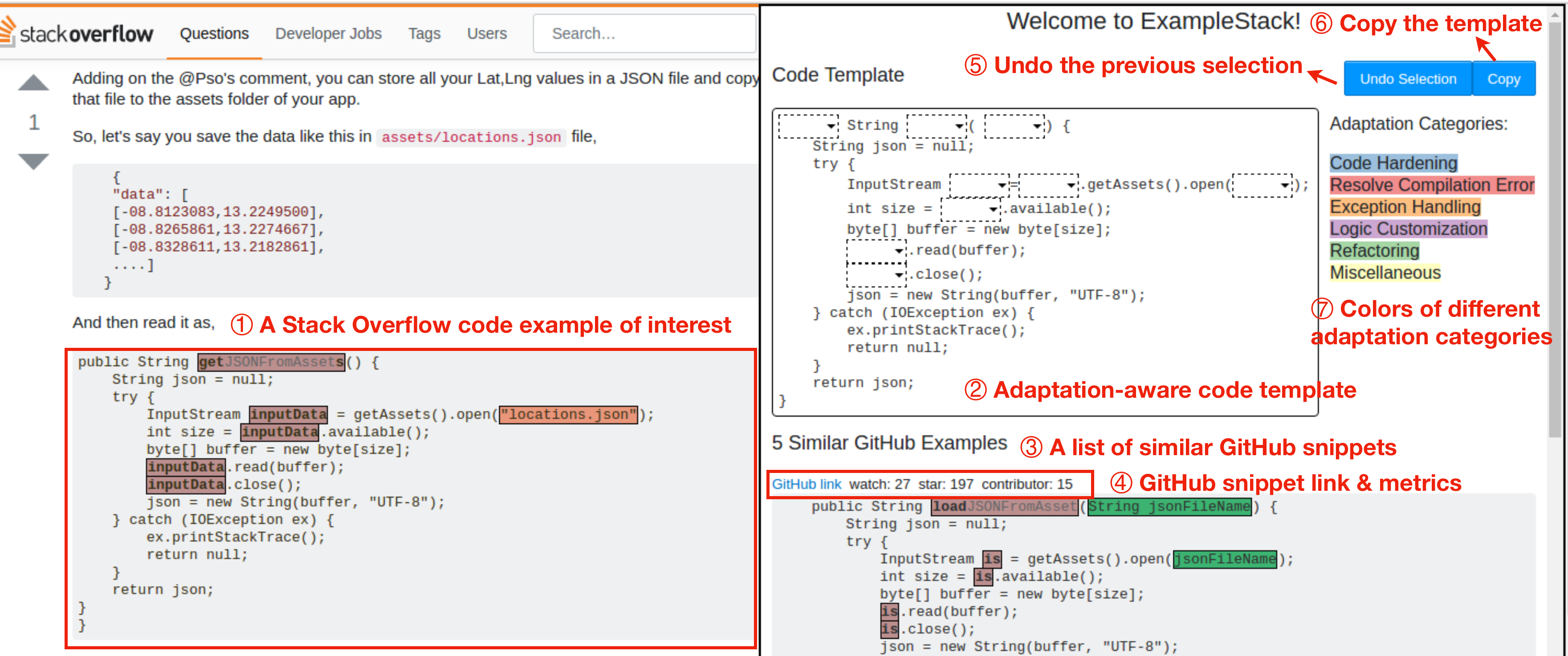}}
\caption{In the lifted template, common unchanged code is retained, while adapted regions are abstracted with {\em hot spots}.}
\label{fig:screenshot}
\end{figure*}

In both datasets, the most frequent adaptation type is {\em change a method call} in the logic customization category. Other logic customization types also occur frequently. This is because SO examples are often designed for illustration purposes with contrived usage scenarios and input data, and thus require further logic customization. {\em Rename} is the second most common adaptation type. It is frequently performed to make variable and method names more readable for the specific context of a GitHub counterpart. 35\% and 14\% of SO examples in the variation dataset and the adaptation dataset respectively include undefined variables or local method calls, leading to compilation errors. The majority of these compilation errors (60\% and 61\% respectively) could be resolved by simply removing the statements using these undefined variables or method calls. 34\% and 22\% of SO examples in the two datasets respectively include new conditionals (e.g., an {\ttt if} check) to handle corner cases or reject invalid input data.

To understand whether the same type of adaptations appears repetitively on the same SO example, we count the number of adaptation types shared by different GitHub counterparts. Multiple clones of the same SO example share at least one same adaptation type in the 70\% of the adaptation dataset and 74\% of the variation dataset. In other words, {\it the same type of adaptations is recurring among different GitHub counterparts}.

\section{Tool Support and Implementation}
\label{sec:tool}
%Motivated by the finding that adaptations and variations of a given SO example often share the same recurring adaptation type, 
Based on insights of the adaptation analysis, we build a Chrome extension called {\tool} that visualizes similar GitHub code fragments alongside a SO code example and allows a user to explore variations of the SO example in an adaptation-aware code template.
%Based on insights of the adaptation analysis, we build a Chrome extension called {\tool} and conduct a within-subjects study with twelve developers to investigate whether developers find it useful to view the commonality and variations between SO code examples and their GitHub counterparts (RQ3). {\tool} visualizes similar GitHub code fragments alongside a SO code example and allows a user to explore potential adaptations in a code template. In the user study, each participant inspects a SO code example and annotates which parts they would like to change during code reuse with or without {\tool}. The result shows that participants give more adaptation suggestions in both number and diversity with {\tool}, taking more usage scenarios and edge cases into account. In the post survey, all developers found {\tool} useful in terms of understanding the essence of a code example and quickly spotting adaptation opportunities in different usage scenarios, especially when browsing a long, difficult code example or when developing a large-scale, robust project. 

\subsection{{\tool} Tool Features}
\label{sec:user}
%This section describes the tool features of {\tool}. 
Suppose Alice is new to Android and she wants to read some {\ttt json} data from the asset folder of her Android application. Alice finds a SO code example~\cite{so-example-json} that reads geometric data from a specific file, {\ttt locations.json} (\ding{172} in Figure~\ref{fig:screenshot}). {\tool} helps Alice by detecting other similar snippets in real-world Android projects and by visualizing the hot spots where adaptations and variations occur. 

{\bf Browse GitHub counterparts with differences.} Given the SO example, {\tool} displays five similar GitHub snippets and highlights their variations to the SO example (\ding{174} in Figure~\ref{fig:screenshot}). It also surfaces the GitHub link and reputation metrics of the GitHub repository, including the number of stars, contributors, and watches (\ding{175} in Figure~\ref{fig:screenshot}). By default, it ranks GitHub counterparts by the number of stars.

\begin{figure}[!h]
\centering
\includegraphics[width=\linewidth]{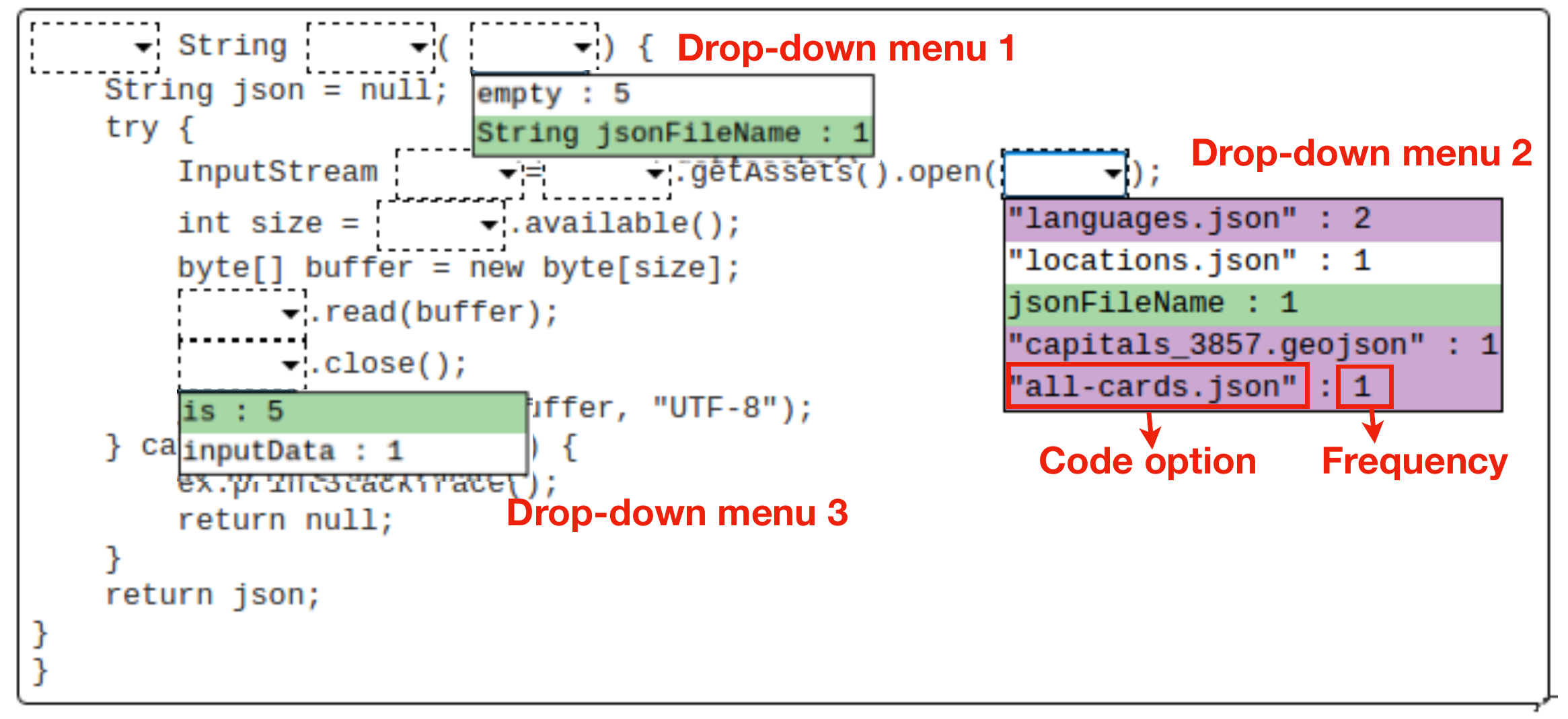}
\caption{Alice can click on a hot spot and view potential code options colored based on their underlying adaptation type.}
\label{fig:dropdown}
\end{figure}

{\bf View hot spots with code options.} {\tool} lifts a code template to illuminate unchanged code parts, while abstracting modified code as {\em hot spots} to be filled in (\ding{173} in Figure~\ref{fig:screenshot}). The lifted template provides a bird's-eye view and serves as a navigation model to explore a variety of code options used to customize the code example. In Figure~\ref{fig:dropdown}, Alice can click on each hot spot and view the code options along with their frequencies in a drop-down menu. Code options are highlighted in six distinct colors according to their underlying adaptation intent (\ding{178} in Figure~\ref{fig:screenshot}). For example, the second drop-down menu in Figure~\ref{fig:dropdown} indicates that two GitHub snippets replace {\ttt locations.json} to {\ttt languages.json} to read the language asset resources for supporting multiple languages. This variation is represented as {\em update a constant value} in the {\em logic customization} category. %in our taxonomy in Section~\ref{sec:categorization}. 

\vspace{-3mm}

\begin{figure}[!h]
\centering
\includegraphics[width=\linewidth]{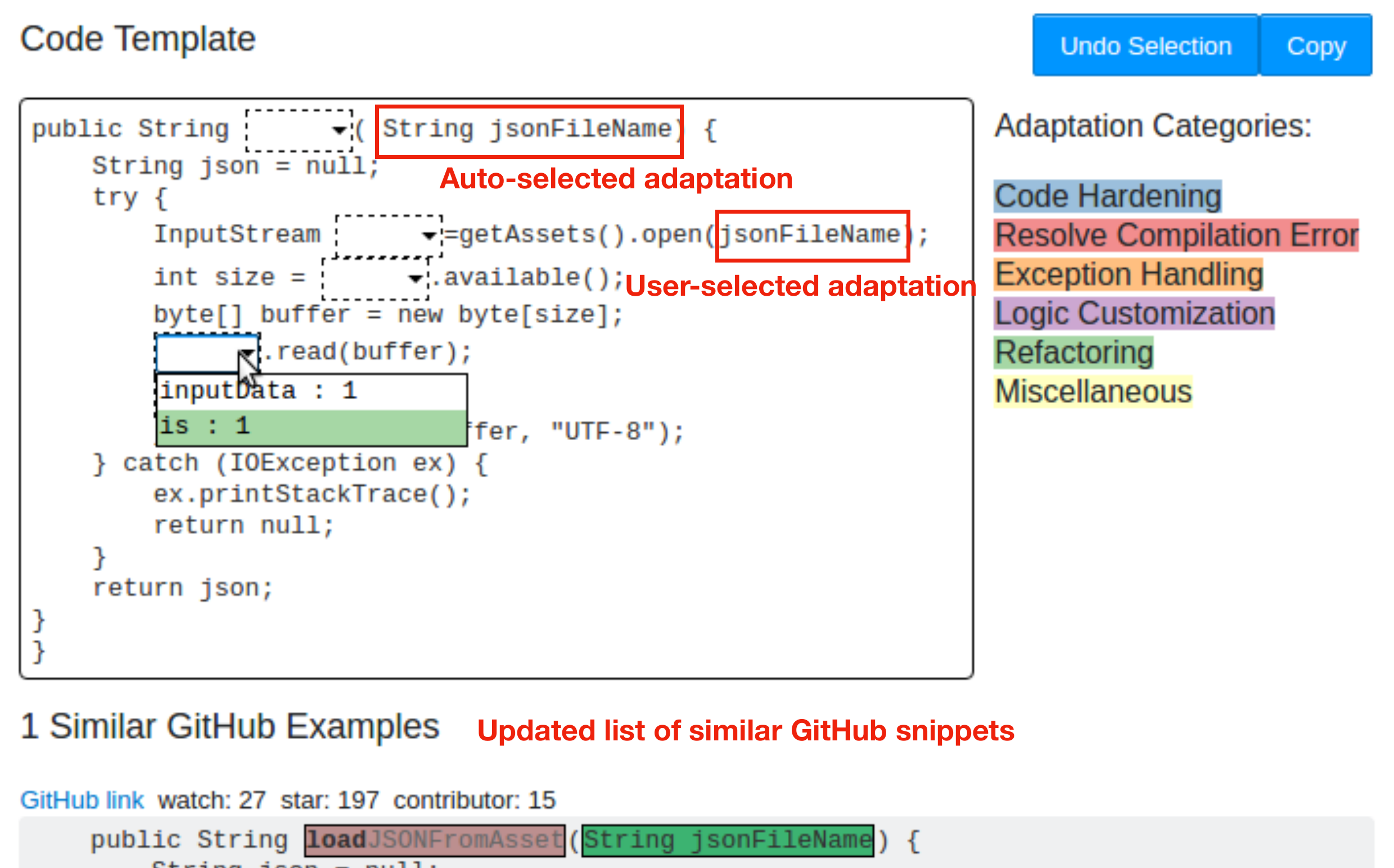}
	\caption{{\tool} automatically updates the code template based on user selection.}  
\label{fig:filter}
\end{figure}

{\bf Fill in hot spots with auto-selection.} Instead of hardcoding the asset file name, Alice wants to make her program more general---being able to read asset files with any given file name. Therefore, Alice selects the code option, {\ttt jsonFileName}, in the second drop-down menu in Figure~\ref{fig:dropdown}, which generalizes the hardcoded file name to a variable. {\tool} automatically selects another code option, {\ttt String jsonFileName}, in the first drop-down menu in Figure~\ref{fig:dropdown}, since this code option declares the {\ttt jsonFileName} variable as the method parameter. This auto-selection feature is enabled by {\it def-use} analysis, which correlates code options based on the definitions and uses of variables (Section~\ref{sec:template}). By automatically relating code options in a template, Alice does not have to manually click through multiple drop-down menus to figure out how to avoid compilation errors. Figure~\ref{fig:filter} shows the customized template based on the selected {\ttt jsonFileName} option. The list of GitHub counterparts and the frequencies of other code options are also updated accordingly based on user selection. Alice can undo the previous selection (\ding{176} in Figure~\ref{fig:screenshot}) or copy the customized template to her clipboard (\ding{177} in Figure~\ref{fig:screenshot}).

\vspace{-3mm}

%\subsection{Multi-program Differencing}
\subsection{Template Construction}
\label{sec:template}

%Figure~\ref{fig:architecture} describes the workflow of how {\tool} lifts a code template, given a SO example and similar GitHub snippets. 

{\bf Diff generating and pruning.} To lift an adaptation-aware code template of a SO code example, {\tool} first computes the AST differences between the SO example and each GitHub clone using GumTree. {\tool} prunes the edit operations by filtering out {\em inner} operations that modify the children of other modified nodes. For example, if an insert operation inserts an AST node whose parent is also inserted by another insert, the first inner insert will be removed, since its edit is entailed by the second outer insert. Given the resulting tree edits, {\tool} keeps track of the change regions in the SO example and how each region is changed. 

%\begin{itemize}[leftmargin=*]
%\item {\bf delete($t$)} is interpreted as {\bf diff<[$start_t$, $end_t$], {\ttt null}>}, where $start_t$ and $end_t$ are the start and end indices of the source code of $t$ in the SO example. Since $t$ is deleted, there is no matched code region in the GitHub snippet.
%\item {\bf insert($t_1$, $t_2$, $i$)} is interpreted as {\bf diff<[$x$,$x$], [$start_{t1}$, $end_{t1}$]>}, where $start_{t1}$ and $end_{t1}$ are the start and end indices of $t_1$ in the GitHub snippet. Though the inserted node, $t_1$, does not have matched code in the SO example, we compute the insertion point, $x$, by matching the surrounding nodes of $t_1$ so that {\tool} later knows where to inject a hole to display the inserted code in the template. 
%\item {\bf move($t_1$, $t_2$, $i$)} is interpreted as {\bf diff<[$start_{t1}$, $end_{t1}$], [$start_{t1}'$, $end_{t1}'$]>}. $start_{t1}$ and $end_{t1}$ are the start and end indices of $t_1$ in the SO example, while $start_{t1}'$ and $end_{t1}'$ are the start and end indices of the matched node of $t_1$ (i.e., the $i$-th child of $t_2$) in the GitHub snippet.
%\item {\bf update($t_1$, $t_2$)} is interpreted as {\bf diff<[$start_{t1}$, $end_{t1}$], [$start_{t2}$, $end_{t2}$]>}. $start_{t1}$ and $end_{t1}$ are the start and end indices of $t_1$ in the SO example, while $start_{t2}$ and $end_{t2}$ are the start and end indices of $t_2$ in the GitHub snippet.
%\end{itemize}

{\bf Diff grouping.} {\tool} groups change regions to decide where to place hot spots in a SO example and what code options to display in a hot spot. If two change regions are the same, they are grouped together. If two change regions overlap, {\tool} merges the overlapping change locations into a bigger region enclosing both and groups them together. For example, consider a diff that changes {\ttt a=\fbox{b}} to {\ttt a=\fbox{b+c}}, and another diff that completely changes \fbox{{\ttt a=b}} to \fbox{{\ttt o.foo()}}. Simply abstracting the changed code in these two diffs without any alignment will overlay two hot spots in the template, \setlength{\fboxsep}{2pt} \fbox{{\ttt a=\fbox{b}}} \setlength{\fboxsep}{3pt} and the smaller diff is shadowed by the bigger diff in visualization. {\tool} avoids this conflict by re-calibrating the first change region from  {\ttt a=\fbox{b}} to \fbox{{\ttt a=b}}.%Since insert operations do not have matched regions in the SO example, {\tool} computes the insertion point of each inserted tree node by matching surrounding tree nodes and clusters the diff regions of insert operations by insertion points.

{\bf Option generating and highlighting.} For each group of change regions, {\tool} replaces the corresponding location in the SO example with a hot spot and attaches a drop-down menu. {\tool} displays both the original content in the SO example and contents of the matched GitHub snippet regions as options in each drop-down menu. {\tool} then uses the adaptation categorization technique to detect the underlying adaptation types of code options. We use six distinct background colors to illuminate the categories in Table~\ref{tab:types}, which makes it easier for developers to recognize different intent. The color scheme is generated using ColorBrewer~\cite{colorbrewer} to ensure the primary visual differences between different categories in the template.

{\tool} successfully lifts code templates in all 14,124 SO examples. On average, a lifted template has 81 lines of code (median 41) with 13 hot spots (median 12) to fill in. On average, 4 code options (median 2) is displayed in the drop-down menu of each hot spot. %Besides the running example in Section~\ref{sec:user}, Figure~\ref{fig:examples} shows three other templates lifted from SO examples and their GitHub counterparts. The template in Figure~\ref{fig:distance} is constructed from one SO example and two GitHub clones about how to compute the distance between two coordinates in terms of longitudes and latitudes. In the template, input types are adapted from {\ttt float} to {\ttt double}, and the value of {\ttt earthRadius} is adapted from {\ttt 637100} meters to {\ttt 3958.75} miles and {\ttt 6371} kilometers to compute distances with different units. A SO user comments that this code example may return negative distances in this post. One GitHub clone~\cite{gh-clone-fix} indeed fixes this bug by adapting the {\ttt return} statement to compute the absolute value of the distance. The template in Figure~\ref{fig:android} illustrates how to add animation to an Android view. The first drop-down menu shows five possible animation types such as fade in (i.e., {\ttt R.layout.activity_fadein}) and blink (i.e., {\ttt R.layout.activity_blink}). One GitHub clone deletes the method call to {\ttt setVisibility}, as indicated by the second drop-down menu. The template in Figure~\ref{fig:hexstring} shows how to encode a byte array filled with hex numbers to its {\ttt string} format. The original method name, {\ttt encodeHexString} is adapted to a variety of names. Two GitHub clones insert a {\ttt null} check before reading data from the input byte array to avoid {\ttt NullPointerException} at runtime.

\begin{table*}[!t]
\caption{Code reuse tasks and user study results}
\label{tab:tasks}
\resizebox{\textwidth}{!}{%
\renewcommand{\arraystretch}{1.3}
\renewcommand\tabcolsep{4pt}
\begin{tabular}{|l|l|r|r|r|l|r|r|l|r|}
\hline
\multirow{2}{*}{{\bf ID}}         & \multirow{2}{*}{{\bf Desired Function \& SO Example}}                                                                                            & \multirow{2}{*}{{\bf LOC}} & \multirow{2}{*}{{\bf Clone\#}} & \multicolumn{3}{c|}{{\bf Control}}                                                                                                                                         & \multicolumn{3}{c|}{{\bf Experiment}}                                                                                                                          \\ \cline{5-10} 
                          &                                                                                                                           &                      &                          & \multicolumn{1}{c|}{Assignment} & \multicolumn{1}{c|}{Adaptation}                                                                    & \multicolumn{1}{c|}{Time(s)} & \multicolumn{1}{c|}{Assignment} & \multicolumn{1}{c|}{Adaptation}                                                        & \multicolumn{1}{c|}{Time(s)} \\ \hline
\multirow{4}{*}{Task I}   & \multirow{4}{*}{\begin{tabular}[c]{@{}l@{}}Calculate the geographic distance \\ between two GPS coordinates~\cite{so-example-distance}\end{tabular}} & \multirow{4}{*}{12}  & \multirow{4}{*}{2}       & P5-A                         & refactor(5), logic(1)                                                                              & 458                           & P2-A                         & harden(1), logic(1), misc(2)                                                           & 870                           \\ \cline{5-10} 
                          &                                                                                                                           &                      &                          & P7-A                         & refactor(1), logic(2), misc(1)                                                                     & 900                           & P3-B                         & refactor(6), logic(4), misc(3)                                                         & 900                           \\ \cline{5-10} 
                          &                                                                                                                           &                      &                          & P12-B                        & refactor(2), harden(1)                                                                             & 900                           & P10-B                        & refactor(5), logic(2), misc(1)                                                         & 366                           \\ \cline{5-10} 
                          &                                                                                                                           &                      &                          & P16-B                        & refactor(7)                                                                                        & 727                           & P15-A                        & refactor(10), logic(14), misc(3)                                                       & 842                           \\ \hline
\multirow{4}{*}{Task II}  & \multirow{4}{*}{\begin{tabular}[c]{@{}l@{}}get the relative path between \\ two files~\cite{so-example-relative-path}\end{tabular}}                       & \multirow{4}{*}{74}  & \multirow{4}{*}{2}       & P3-A                         & refactor(5), logic(1), exception(2), misc(3)                                                       & 900                           & P1-B                         & refactor(3), harden(1), logic(2)                                                       & 640                           \\ \cline{5-10} 
                          &                                                                                                                           &                      &                          & P8-A                         & harden(1)                                                                                          & 900                           & P6-A                         & harden(4), logic(3)                                                                    & 900                           \\ \cline{5-10} 
                          &                                                                                                                           &                      &                          & P11-B                        & none                                                                                               & 621                           & P9-A                         & harden(4), logic(2)                                                                    & 900                           \\ \cline{5-10} 
                          &                                                                                                                           &                      &                          & P15-B                        & \begin{tabular}[c]{@{}l@{}}refactor(13), harden(1), logic(5),\\ exception(1), misc(1)\end{tabular} & 863                           & P13-B                        & \begin{tabular}[c]{@{}l@{}}refactor(3), logic(2), exception(1),\\ misc(1)\end{tabular} & 900                           \\ \hline
\multirow{4}{*}{Task III} & \multirow{4}{*}{\begin{tabular}[c]{@{}l@{}}encode a byte array to a \\ hexadecimal string~\cite{so-example-encode}\end{tabular}}                   & \multirow{4}{*}{12}  & \multirow{4}{*}{17}      & P1-A                         & refactor(5), harden(1)                                                                             & 652                           & P4-A                         & refactor(5), harden(1), misc(1)                                                        & 667                           \\ \cline{5-10} 
                          &                                                                                                                           &                      &                          & P6-B                         & refactor(1), misc(1)                                                                               & 900                           & P8-B                         & refactor(2), harden(1), misc(2)                                                        & 548                           \\ \cline{5-10} 
                          &                                                                                                                           &                      &                          & P9-B                         & harden(1), logic(1)                                                                                & 635                           & P12-A                        & refactor(3), harden(2), misc(1)                                                        & 748                           \\ \cline{5-10} 
                          &                                                                                                                           &                      &                          & P13-A                        & refactor(3), misc(1)                                                                               & 900                           & P14-B                        & refactor(3), harden(1), misc(1)                                                        & 700                           \\ \hline
\multirow{4}{*}{Task IV}  & \multirow{4}{*}{\begin{tabular}[c]{@{}l@{}}add animation to an Android \\ view~\cite{so-example-android}\end{tabular}}                              & \multirow{4}{*}{29}  & \multirow{4}{*}{4}       & P2-B                         & refactor(3), logic(1)                                                                              & 441                           & P5-B                         & refactor(1), logic(3)                                                                  & 478                           \\ \cline{5-10} 
                          &                                                                                                                           &                      &                          & P4-B                         & refactor(1), compile(1), misc(1)                                                                   & 900                           & P7-B                         & refactor(2), compile(3), logic(3)                                                      & 887                           \\ \cline{5-10} 
                          &                                                                                                                           &                      &                          & P10-A                        & refactor(3), logic(5)                                                                              & 900                           & P11-A                        & refactor(1), logic(3)                                                                  & 617                           \\ \cline{5-10} 
                          &                                                                                                                           &                      &                          & P14-A                        & refactor(2), logic(4)                                                                              & 862                           & P16-A                        & refactor(6), logic(4), misc(1)                                                         & 773                           \\ \hline
\end{tabular}
}%
\end{table*}

\section{User Study}
\label{sec:user-study}

We conducted a within-subjects user study with sixteen Java programmers to evaluate the usefulness of %viewing similar GitHub code alongside SO examples using
{\tool}. 
%The goal was to investigate whether seeing commonalities and variations in similar GitHub code could inspire developers with new adaptations that they may otherwise ignore. 
We emailed students in a graduate-level Software Engineering class and research labs in the CS department at UCLA. We did a background survey and excluded volunteers with no Java experience, since our study tasks required users to read code examples in Java. Fourteen participants were graduate students and two were undergraduate students. 
%Since our study tasks required participants to read code examples in Java, we only included students who had taken at least one Java class. 
Eleven participants had two to five years of Java experience, while the other five were novice programmers with one-year Java experience, showing a good mix of different levels of Java programming experience. % All participants reported to use Stack Overflow for programming. Seven participants said they visited Stack Overflow everyday, four visited at least once a week, and one visited once every few months. All but one partcipants reported that they often reused and adapted code examples from Stack Overflow. 

In each study session, we first gave a fifteen-minute tutorial of our tool. Participants then did two code reuse tasks with and without {\tool}. When not using our tool (i.e., the control condition), participants were allowed to search online for other code examples, which is commonly done in real-world programming workflow~\cite{brandt2009two}. To mitigate learning effects, the order of assigned conditions and tasks were counterbalanced across participants through random assignment. In each task, we asked participants to mark which parts of a SO code example they would like to change and explain how they would change. We did not require participants to fully integrate a code example to a target program or make it compile, since our goal was to investigate whether {\tool} could inspire developers with new adaptations that they may otherwise ignore, rather than automated code integration. Each task was stopped after fifteen minutes. At the end, we did a post survey to solicit feedback.

\begin{figure*}[!t]
\begin{minipage}[c]{\linewidth}
  \begin{subfigure}[b]{0.36\linewidth}
    \includegraphics[width=\linewidth]{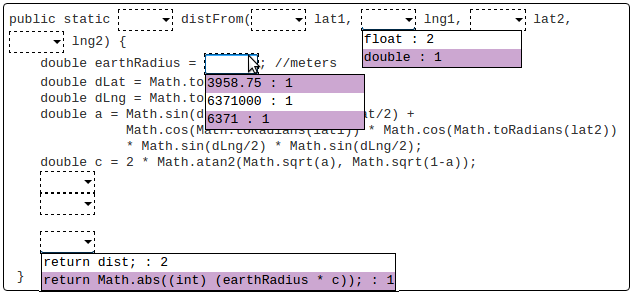}
    \caption{Compute distance between two coordinates~\cite{so-example-distance}}
    \label{fig:distance}
  \end{subfigure}
  \begin{subfigure}[b]{0.32\linewidth}
    \includegraphics[width=\linewidth]{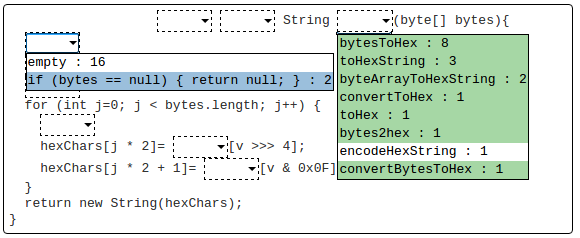}
    \caption{Encode byte array to a hex string~\cite{so-example-encode}}
    \label{fig:hexstring}
  \end{subfigure} 
  \begin{subfigure}[b]{0.32\linewidth}
    \includegraphics[width=\linewidth]{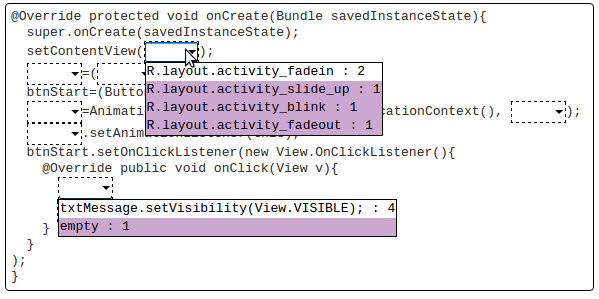}
    \caption{Add animation to an Android view~\cite{so-example-android}}
    \label{fig:android}
  \end{subfigure}
  \end{minipage}
  \caption{{\tool} code template examples}
  \label{fig:examples}
\end{figure*}

Table~\ref{tab:tasks} describes the four code reuse tasks and also the user study results. Column {\sf Assignment} in each condition shows the participant ID and the task order. ``P5-A'' means the task was done by the fifth participant as her first task. Column {\sf Adaptation} shows the number of different types of adaptations each participant made. Overall, participants using {\tool} made three times more code hardening adaptations (15 vs.~5) and twice more logic customization adaptations (43 vs.~20), considering more edge cases and different usage scenarios. For instance, in Task III, all users in the experimental group added a null check for the input byte array after seeing other GitHub examples, while only one user in the control group did so. P14 wrote, ``{\em I would have completely forgotten about the null check without seeing it in a couple of examples.}'' On average, participants using {\tool} made more adaptations (8.0 vs. 5.5) in more diverse categories (2.8 vs. 2.2). Wilcoxon signed-rank tests indicate that the mean differences in adaptation numbers and categories are both statistically significant (p=0.042 and p=0.009). We do not argue that making more adaptations are always better. Instead, we want to emphasize that, by seeing commonalities and variations in similar GitHub code, participants focus more on code safety and logic customization, instead of making shallow adaptations such as variable renaming only. The average task completion time is 725 seconds (SD=186) and 770 seconds (SD=185) with and without {\tool}. We do not claim {\tool} saves code reuse time, since it is designed as an informative tool when developers browse online code examples, rather than providing direct code integration support in an IDE. Figure~\ref{fig:examples} shows the code templates generated by {\tool}, not including the one in Task II due to its length (79 lines).

{\bf\em How do you like or dislike viewing similar GitHub code alongside a SO example?} In the post survey, all participants found it very useful to see similar GitHub code for three main reasons. First, viewing the commonality among similar code examples helped users quickly understand the essence of a code example. P6 described this as ``{\em the fast path to reach consensus on a particular operation.}'' Second, the GitHub variants reminded users of some points they may otherwise miss. 
%P2 wrote, ``{\em I learned different ways to implement something and understand the nuance better by seeing what is and is not similar.}'' 
Third, participants felt more confident of a SO example after seeing how similar code was used in GitHub repositories. P9 stated that, ``{\em [it is] reassuring to know that the same code is used in production systems and to know the common pitfalls.}'' 

{\bf\em How do you like or dislike interacting with a code template?} Participants liked the code template, since it showed the essence of a code example and made it easier to see subtle changes, especially in lengthy code examples. Participants also found displaying the frequency count of different adaptations very useful. P5 explained, ``{\em it highlights the best practices followed by the community and also prioritizes the changes that I should make first.}'' However, we also observed that, when there were only a few GitHub counterparts, some participants inspected individual GitHub counterparts directly rather than interacting with the code template. 

{\bf\em How do you like or dislike color-coding different adaptation types?} Though the majority of participants confirmed the usefulness of this feature, six participants felt confused or distracted by the color scheme, since it was difficult to remember these colors during navigation. Three of them considered some adaptations (e.g., renaming) trivial and suggested to allow users to hide adaptations of no interest to avoid distraction.

{\bf\em When would you use {\tool}?} Six participants would like to use {\tool} when learning APIs, since it provided multiple GitHub code fragments that use the same API in different contexts with critical safety checks and exception handling. Five participants mentioned that {\tool} would be most useful for a lengthy example. 
%P8 said, ``{\em I would use this tool when I am not sure of the credibility of the SO code.}'' 
P4 wrote, ``{\em the tool is very useful when the code is longer and hard to spot what to change at a glance.}''  Two participants wanted to use {\tool} to identify missing points and assess different solutions, when writing a large-scale robust project.

%{\bf\em How would you like to improve our tool?} In addition to making the color scheme customizable and hiding certain adaptation types, multiple participants suggested a meta-voting feature that allows users to rate the usefulness of individual GitHub examples and adaptations in {\tool}. 
In addition, P15 and P16 suggested to display similar code based on semantic similarity rather than just syntactic similarity, in order to find alternative implementations and potential optimization opportunities. P13 suggested to add indicators about whether a SO example is compilable or not. 

\section{Threats to Validity}
\label{sec:discussion}

%\noindent{\bf\em Threats to Validity.}
In terms of {\em internal validity}, our variation dataset may include coincidental clones, since GitHub developers may write code with similar functionality as a SO example. To mitigate this issue, we compare their timestamps and remove those GitHub clones that are created before the corresponding SO examples. We further create an adaptation set with explicitly attributed SO examples and compare the analysis results of both datasets for cross-validation. Figure~\ref{fig:adapt-type-frequency} shows that the distribution of common adaptation patterns is similar between these two datasets. It would be valuable and useful to guide code adaptation by identifying the commonalities and variations between similar code, even for clones coming from independent but similar implementations.
%In the dataset construction, identifying which SO example is attributed only based on SO links in a GitHub file may produce mismatches, if there are multiple similar examples in an answer post or a discussion thread. Clone detection might introduce false positive clones that are only structurally similar. 
%We have made our best effort in conducting a comprehensive and unbiased dataset via extra quality control mechanisms such as time order checking, manual inspection, and cross-validation.

In terms of {\em external validity}, when identifying common adaptation types, we follow the standard qualitative analysis procedure~\cite{berg2004qualitative} to continuously inspect more samples till the insights are converging. However, we may still miss some adaptation types due to the small sample size. To mitigate this issue, the second author who was not involved in the manual inspection further manually labeled 100 more samples to validate the adaptation taxonomy (Section~\ref{sec:eval1}). In addition, user study participants may not be representative of real Stack Overflow users. To mitigate this issue, we recruit both novice and experienced developers who use Stack Overflow on a regular basis. To generalize our findings to industrial settings, further studies with professional developers are needed.

In terms of {\em construct validity}, in the user study, we only measure whether {\tool} inspires participants to identify and describe adaptation opportunities. We do not ask participants to fully integrate a SO example to a target program nor make it compile. Therefore, our finding does not imply time reduction in code integration. %Since the user study is designed to understand whether developers can grasp a comprehensive view of adapting a SO example using {\tool}, we do not ask users to perform code integration to avoid confounding factors such as the familiarity of IDEs and the complexity of target programs. Instead, we ask users to annotate where and how they would like to change a SO example and encourage them to think about possible adaptations in different usage cases. Therefore, our findings do not imply code integration effectiveness.

\section{Related Work}
\label{sec:relwork}
\noindent{\bf\em Quality assessment of SO examples.} Our work is inspired by previous studies that find SO examples are incomplete and inadequate~\cite{dagenais2012recovering, subramanian2013making, yang2016query, zhou2016api, fischer2017stack, an2017stack, icse2018}. %Yang et al.~analyze 3 million SO code snippets and find that code in Java and C\# cannot be parsed or compiled by standard compilers~\cite{yang2016query}. Dagenais and Robillard find that 89\% of API names in code snippets from online forums are ambiguous and cannot be easily resolved due to incompleteness~\cite{dagenais2012recovering}. 
Subramanian and Holmes find that the majority of SO snippets are free standing statements with no class or method headers~\cite{subramanian2013making}. 
Zhou et al.~find that 86 of 200 accepted SO posts use deprecated APIs but only 3 of them are reported by other programmers~\cite{zhou2016api}. Fischer et al.~find that 29\% of security-related code in SO is insecure and could potentially be copied to one million Android apps~\cite{fischer2017stack}. %Treude and Robillard conduct a survey to investigate comprehension difficulty of code examples in SO~\cite{treude2017understanding}. The responses from GitHub users indicate that less than half of the SO examples are self-explanatory and one main issue is code incompleteness. 
Zhang et al.~contrast SO examples with API usage patterns mined from GitHub and detect potential API misuse in 31\% of SO posts~\cite{icse2018}. These findings motivate our investigation of adaptations and variations of SO examples. 

\noindent{\bf\em Stack Overflow usage and attribution.} 
Our work is motivated by the finding that developers often resort to online Q\&A forums such as Stack Overflow~\cite{sadowski2015developers, brandt2009two, wu2018developers, baltes2018usage}. Despite the wide usage of SO, most developers are not aware of the SO licensing terms nor attribute to the code reused from SO~\cite{an2017stack, baltes2018usage, wu2018developers}. Only 1.8\% of GitHub repositories containing code from SO follow the licensing policy properly~\cite{baltes2018usage}. Almost one half developers admit copying code from SO without attribution and two thirds are not aware of the SO licensing implications. Based on these findings, we carefully construct a comprehensive dataset of reused code, including both explicitly attributed SO examples and potentially reused ones using clone detection, timestamp analysis, and URL references. 
%A case study at Google shows that developers issue an average of 12 code search queries per weekday~\cite{sadowski2015developers}. 
%Brandt et al.~find that developers often use online resources to learn new concepts and remind themselves about programming details~\cite{brandt2009two}.
Origin analysis can also be applied to match SO snippets with GitHub files~\cite{tu2002integrated, godfrey2002tracking, zou2003detecting, godfrey2005using}. %Unlike clone detection, origin analysis leverages the LCS similarity between program entities such as variable names and expressions as well as call relations to identify the origin of a piece of code during software evolution.

%An et al.~investigate copyright issues between SO and GitHub~\cite{an2017stack} and find a large number of potential license violations. 
%Yang et al.~run clone detection between 1.9M python snippets on SO and 909K non-fork python projects hosted on GitHub~\cite{yang2017stack}. They find 48K python snippets have counterpart GitHub clones, which could be potentially copied and pasted to GitHub. Our work differs from these studies by focusing on analysis of SO example adaptations. %Our insight is that, by displaying similar code in real-world projects with their variations, developers can better understand limitations of curated examples and recognize what other developers often change when reusing the same example to real software systems.

\noindent{\bf\em SO snippet retrieval and code integration.} Previous support for reusing code from SO mostly focuses on helping developers locate relevant posts or snippets from the IDE~\cite{bacchelli2012harnessing, ponzanelli2013seahawk, ponzanelli2014mining, wightman2012snipmatch}. For example, Prompter retrieves related SO discussions based on the program context in Eclipse. SnipMatch supports light-weight code integration by renaming variables in a SO snippet based on corresponding variables in a target program~\cite{wightman2012snipmatch}. Code correspondence techniques~\cite{holmes2007supporting, cottrell2008semi} match code elements (e.g., variables, methods) to decide which code to copy, rename, or delete during copying and pasting. Our work differs by focusing on analysis of common adaptations and variations of SO examples. 

%Though curated examples can serve a good starting point, they could potentially impact the quality of production code, when integrated to a target application verbatim. Therefore, it is beneficial to visualize the commonality and variations between a curated example and similar code in GitHub so that developers can have a better understanding about what adaptations are needed when reusing a curated example.
%the main issues include incomplete code, code quality, missing rationale, code organization, clutter, naming issues, and missing domain information. 

%Prior research in software engineering has studied code cloning within software projects through both automated~\cite{Baxter1998} and ethnographic~\cite{Kim04} approaches. \todo{add more studies about code clones and also add related work about clone detection.} Kim et al.~find that intra-project clones were unstable and often led to inconsistencies between clones~\cite{Kim04}. However, prior work has been focused on studying intra-project clones and minimizing intra-project duplicated code to reduce maintenance costs~\cite{hua2015does}. %Our study shows that the conventional wisdom about intra-project clones may not apply to online code reuse and should not obviate the potential value of reusing mature and well-recognized code snippets from the Internet.

%\noindent{\bf\em Code Search and Integration.} Seahawk (code search), Jigsaw (code integration), Codota/Examplore (API usage) No one aims to show a bunch of similar examples and demonstrate their variations.

\noindent{\bf\em Change types and taxonomy.} %According to ISO/IEC 14764~\cite{iso}, there are four general categories of program changes---{\em corrective}, {\em adaptive}, {\em perfective}, and {\em preventive}. 
There is a large body of literature for source code changes during software evolution~\cite{Kim2006:MBF, Fischer2005, dig2006apis}.  
Fluri et al.~present a fine-grained taxonomy of code changes such as changing the return type and renaming a field, based on differences in abstract syntax trees~\cite{Fluri2006}. Kim et al.~analyze changes on ``micro patterns''~\cite{gil2005micro} in Java using software evolution data~\cite{kim2006micro}. These studies investigate general change types in software evolution, while we quantify common adaptation and variation types using SO and GitHub code.

%However, this taxonomy is designed purely based on edits in abstract syntax trees and only support basic edit operations, including insert, delete, move, and update. Such syntatic change taxonomy often lacks insights of each change type and also does not support composite change types beyond basic edit operations. To understand the adaptations required to reuse a curated example to a target application is under-investigated, we conduct a large-scale analysis on the clone variations between SO snippets and their GitHub counterparts. Based on the analysis results and insights from the real-world dataset, we design the first change taxonomy that quantifies the necessary adaptations when reusing online code examples and provide tool support to automatically detect and highlight important adapations.

\noindent{\bf\em Program differencing and change template.} Diff tools compute program differences between two programs~\cite{miller1985file, Yang1991, Chawathe1996, fluri2007change, falleri2014fine}. %For example, {\em diff} computes line-level differences using the longest common subsequence algorithm. ChangeDistiller builds mappings and generates edit scripts between two ASTs~\cite{fluri2007change}. GumTree computes differences inside a statement and detects move operations more accurately than ChangeDistiller~\cite{falleri2014fine}. 
However, they do not support analysis of one example with respect to multiple counterparts simultaneously. Lin et al.~align multiple programs and visualize their variations~\cite{lin2014detecting}. However, they do not lift a code template to summarize the commonalities and variations between similar code. Several techniques construct code templates for the purpose of code search~\cite{zhang2015interactive} or code transformation~\cite{meng2013lase}. %Lase takes multiple change examples as input and generalizes the differences for the purpose of applying similar edits~\cite{meng2013lase}. Critics requires users to iteratively parameterize a change template for the purpose of searching similar changes~\cite{zhang2015interactive,zhang2014critics}. 
%{\tool} differs from these by identifying GitHub snippets similar to SO examples, annotating a lifted template with common adaptation types, and visualizing concrete variations in {\it hot spots}. 
Glassman et al.~design an interactive visualization called {\examplore} to help developers comprehend hundreds of similar but different API usages in GitHub~\cite{chi2018}. Given an API method of interest, {\examplore} instantiates a pre-defined API usage skeleton and fills in details such as various guard conditions and succeeding API calls. {\tool} is not limited to API usage and does not require a pre-defined skeleton.

\section{Conclusion}
\label{sec:conclusion}

This paper provides a comprehensive analysis of common adaptation and variation patterns of online code examples by both overapproximating and underapproximating reused code from Stack Overflow to GitHub. Our quantitative analysis shows that the same type of adaptations and variations appears repetitively among different GitHub clones of the same SO example, and variation patterns resemble adaptation patterns. This implies that different GitHub developers may apply similar adaptations to the same example over and over again independently. This further motivates the design of {\tool}, a Chrome extension that guides developers in adapting online code examples by unveiling the commonalities and variations of similar past adaptations. A user study with sixteen developers demonstrates that {\tool} helps developers focus more on code safety and logic customization during code reuse, resulting in more complete and robust code.

%This paper reports common adaptations and variations of online code examples along with their types and frequencies using large-scale Stack Overflow and GitHub repository data. We construct a taxonomy of 24 common adaptation types through manual inspection and further develop an automated technique to categorize AST edits to these adaptation categories. Our automated adaptation categorization technique has 98\% precision and 96\% recall on the manually validated ground truth. By applying this automated analysis to the carefully constructed datasets of {\em adaptations} and {\em variations}, we find that the same type of adaptation appears repetitively between SO examples and GitHub counterparts, implying that developers should benefit from being informed of previous adaptations and variations. We design and implement a Chrome extension called {\tool} to render the commonality and differences between SO examples and similar GitHub snippets by lifting an adaptation-aware code template. A user study with sixteen participants demonstrates that seeing commonalities and variations iof similar GitHub code examples inspires developers to consider different reuse scenarios and edge cases, resulting in more complete and robust code.
%crowd-sourced insights from prior adaptations and variations help improve confidence about how to reuse online code examples.  

%\noindent{\bf\em Future Work.} 
Currently, {\tool} only visualizes potential adaptations of a SO example within a web browser. As future work, we plan to build an Eclipse plugin that enables semi-automated integration of online code examples. It would be worthwhile to investigate how such a tool fits developer workflow and to compare it with other code integration techniques~\cite{cottrell2008semi, wightman2012snipmatch}.
%Existing code integration techniques can automatically rename variables and port related program statements based on the context of a target program during code reuse~\cite{cottrell2008semi, meng2013lase, wightman2012snipmatch}. This paper highlights opportunities for extending such techniques. For example, commonalities shown in type conversion may suggest which type to convert which. 
%Furthermore, developers extensively refactor and add code comments to improve the readability and maintainability of the adapted code. Therefore, automated name synthesis and comment generation could be beneficial. 
%We also plan to leverage more semantic relationships to associate code options in different drop-down menus for auto-selection. For example, we can use mined API usage patterns to relate code options of API methods that are used together, such as {\ttt lock} and {\ttt unlock}.

\section*{Acknowledgment}
Thanks to anonymous participants for the user study and anonymous reviewers for their valuable feedback. This work is supported by NSF grants CCF-1764077, CCF-1527923, CCF-1460325, CCF-1723773, ONR grant N00014-18-1-2037, Intel CAPA grant, and DARPA MUSE program.

\bibliographystyle{IEEEtran}
\bibliography{draft}

\end{document}